\documentclass[]{jpsj2}
\usepackage{graphicx}
\usepackage{latexsym}
\usepackage{amsmath}
\usepackage{color}

\def\dis{\displaystyle}

\title{%
Stochastic Approach to Enantiomeric Excess Amplification and Chiral Symmetry Breaking
}

\author{%
Yukio \textsc{Saito}\thanks{yukio@rk.phys.keio.ac.jp},
Takeshi \textsc{Sugimori}\thanks{tsugimor@rk.phys.keio.ac.jp}, 
 and Hiroyuki \textsc{Hyuga}\thanks{hyuga@rk.phys.keio.ac.jp}
}

\inst{%
Department of Physics, Keio University, Yokohama 223-8522 
}

\recdate{\today}

\abst{%
Stochastic aspects of 
chemical reaction models related to the Soai reactions as well as to the homochirality in life
 are studied analytically and numerically by the use of the master equation
 and random walk model.
For systems with a recycling process, a unique 
final probability distribution is obtained by means of detailed balance
conditions.
With a nonlinear autocatalysis the distribution has
a double-peak structure,
indicating the chiral symmetry breaking. 
This problem is further analyzed by examining eigenvalues and eigenfunctions
of the master equation. 
In the case without recycling process,
 final probability distributions depend
on the initial conditions. In the nonlinear autocatalytic case,
 time-evolution starting from a complete achiral state leads
to a final distribution which differs from that deduced from the nonzero
recycling result. This is due to the absence of the detailed balance,
 and a directed random walk model is shown to give the correct final profile.
When the nonlinear autocatalysis is sufficiently strong and the initial state is
achiral,
the final probability distribution has a double-peak structure,
related to the enantiomeric excess amplification. 
It is argued that with autocatalyses
 and a very small but nonzero spontaneous production, a single mother scenario
 could be a main mechanism to produce the homochirality.
 }

\kword{%
Soai reaction, homochirality,
chirality selection, enantiometric excess, 
probability distribution, master equation,
nonlinear autocatalysis, recycling, directed random walk
}

\begin{document}
\sloppy
\maketitle

\section{Introduction}

For some organic molecules two kind of stereostructures that are mutually 
mirror symmetric are possible to exist, and they are called enantiomers
\cite{vollhardt+99}.
Two enantiomers are chiral like the left- and the right-hand which
cannot be overlapped by translational and rotational transformations.
Since their physical properties are mostly the same except the
response to the optical polarity, there should be an equal amount
of both enantiomers in the production started from achiral substrates.
In nature, on the other hand, it has long been known that the chiral symmetry in life is broken
\cite{pasteur849,japp898};
all proteins consist of the left-handed {\small L}-amino acids 
and nearly all sugars belong to the right-handed {\small D}-series
\cite{stryer98}.
The origin of this homochirality has attracted much attention
in relation to the origin of life itself 
\cite{japp898,vollhardt+99,calvin69,goldanskii+88,gridnev06}
Ideas explaining the origin of homochirality are categorized in two groups;
it is brought by some external advantage factor for one chirality to the other
\cite{japp898},
or it happens by accident \cite{peason898}.
In both cases, however, the degree of excess of one enantiomer to the other is expected
very small \cite{mills32,goldanskii+88}, and the amplification of 
enantiomeric excess (ee)
is indispensable. 

In an open system, Frank proposed long ago a theoretical model 
which contains autocatalysis and an antagonistic nonlinear process, 
and he showed that the model leads
to a chirality selection \cite{frank53}.
Recently Soai and his coworkers found experimental systems
 which show the ee amplification \cite{soai+95,soai+00}.
Experiments were performed in a closed system, and the ee
amplification was shown to depend on the initial condition.
This result is explained by assuming the quadratic autocatalysis \cite{sato+01,sato+03}.
By including a recycling process in addition, it is shown that the system relaxes
to a unique final state with a broken chiral symmetry \cite{saito+04a}.

Most of the theoretical analyses have been performed 
in terms of deterministic rate equations
\cite{saito+04a,saito+04b,saito+05a,saito+05b,saito+05c,shibata+06,
saito+07,sato+01,sato+03,blackmond+01,buhse03,islas+05}.
In the meanwhile, sequences of experimental runs 
starting from an initial state with no chiral ingredients have been performed 
\cite{soai+03,singleton+02,gridnev+03,singleton+03}. 
In some runs, the system has a preference to one chirality, and in
some other runs to the other chirality. 
Values of an ee order parameter are distributed wide in a whole possible range.
The probability distribution for the ee value has symmetric double peaks 
at intermediate values of ee with opposite signs.
Symmetric profile reflects the fact that there is no preference 
in the chirality in the initial situation, and
the double-peak structure represents that the symmetry breaking is
induced in the dynamic evolution.
For the description of the probability distribution of populations of chiral species, we have to study the
stochastic aspects of the chiral molecule production.
There is a stochastic study of a spontaneous and a linear autocatalytic
chemical reaction without recycling \cite{lente04,lente05}.
Here we study stochastic evolution of systems with 
more generic chemical reactions;
linear as well as nonlinear autocatalysis with and without recycling.

In \S 2, we present a stochastic model for the production of chiral molecules
$R$ and $S$ from an achiral substrate $A$ in a closed system. 
Transition probabilities contain processes as
a spontaneous production,
a linear and a quadratic autocatalytic production, and
a  back reaction which recycles the substrate from the products.
In \S 3, we discuss the shape of the final probability distribution
by assuming a detailed balance between production and recycling reactions. 
This analysis is valid 
if the final probability distribution is unique, independent of the intermediate
path. The peak position of the final probability distribution is found to
correspond to the fixed points of the rate equations.
With a quadratic autocatalysis, 
the distribution has a double-peak profile,
indicating the chiral symmetry breaking.
In \S 4 the master equation which governs the time evolution of 
the probability distribution is integrated numerically, and
the final distribution is compared with that obtained in the previous section.
In most cases two results agree, but
if the system has a quadratic autocatalysis without recycling back reaction,
the numerical integration gives rise to a  profile different from that expected  
in the limit of weak recycling.
The chiral symmetry breaking in the case with recycling is interpreted
in terms of a degeneracy in the eigenvalues of the master equation
evolution in \S 5. 
The zero eigenvalue state corresponds to the final probability distribution,
and with a quadratic autocatalysis the first non-zero 
eigenvalue approaches zero
in the large number limit of the reactants.
Without recycling, there are infinitely many degeneracy of eigenvalues
at zero,
representing the non-ergodicity such that the final probability 
depends on the initial condition.
In \S 6 a toy model for the system without recycling, 
a directed random walk model, is proposed.
It gives the final probability distribution for a spontaneous and a linear
autocatalytic system analytically, and for a quadratic
autocatalytic system numerically. 
The resulting distributions agree with those given by the numerical integration
of the master equation in \S 4.
The stochastic approach reveals (i) the single mother scenario of the
homochirality:
In the case of a very small spontaneous production rate, 
the first mother chiral species produced by this spontaneous production 
converts all the substrate molecules into her type of chiral products by some
autocatalytic processes, 
before the second mother of another chiral type is born.
(ii) With a quadratic autocatalysis with a moderate value of the rate constant,
double peaks appear in a probability distribution
at a finite values of the ee order parameter,
in agreement with the experimental observation:
The spontaneously produced racemic single peak splits due to the
nonlinear amplification of chiral imbalance.
The result is summarized in the last section.

\section{Model and Elementary Processes}
In order to understand the stochastic features of chiral symmetry breaking,
we use here the simplest model considered previously
\cite{saito+04a}:
An achiral substrate molecule $A$
turns into one of the two enantiomers $R$ or $S$ in a closed system.
Only with spontaneous reactions 
\begin{align}
A \rightarrow R, \qquad A \rightarrow S
\label{eq01}
\end{align}
with the same reaction rate $k_0$, it is easily shown that an
initial enantiomeric excess (ee) decreases
\cite{saito+04a}.
For the ee amplification, some autocatalytic processes are necessary. 
We consider two types of autocatalyses; a linear and a quadratic ones.
A rate constant of linear autocatalyses 
\begin{align}
A +R \rightarrow 2R, \qquad A+S \rightarrow 2S
\label{eq02}
\end{align}
is set $k_1$, and that of quadratic autocatalyses 
\begin{align}
A + 2R \rightarrow 3R, \qquad A+2S \rightarrow 3S
\label{eq03}
\end{align}
 is set $k_2$.
It is found \cite{saito+04a}
that additional back reactions from the chiral products 
$R$ or $S$ to the achiral substrate $A$, 
which we call recycling processes hereafter,
\begin{align}
R \rightarrow A, \qquad S \rightarrow A
\label{eq04}
\end{align}
select a unique final state with a definite value of the ee:
The situation is similar to the symmetry breaking in equilibrium
statistical physics, and we call it a chiral symmetry breaking.
By denoting the rate of the recycling (\ref{eq04}) as $\lambda$,
the rate equations of the concentrations of chiral enantiomers
$r$ and $s$ are written as
\begin{align}
\frac{dr}{dt}  = (k_0+k_1r+k_2r^2) a - \lambda r,
\nonumber \\
\frac{dr}{dt}  = (k_0+k_1s+k_2s^2) a - \lambda s.
\label{eq05}
\end{align}
Because the system is assumed to be closed, the total concentration $c$ of all the 
reactant species is conserved
and the concentration of achiral substrate $a$
is determined as $a=c-r-s$. The order parameter of ee is defined as usual as
\begin{align}
\phi=\frac{r-s}{r+s}
\label{eq05a}
\end{align}
and its absolute value $|\phi|$ is often called an ee parameter.

Rate equations in general describe {\it averaged} behaviors of the
reaction system when there are ample amount of molecules.
In the initial stage of reaction, however, there are only a very few
product molecules $R$ and $S$ and the discrete and the stochastic aspect 
of the chemical reaction 
could be important for autocatalytic reactions. 
The state of a system is described by population numbers of achiral and
chiral molecules as $N_A,~N_R,~N_S$ where the total number
of molecules $N$ are fixed constant so that $N_A=N-N_R-N_S$.
At a time $t$, the system is in a state $(N_A, N_R,~N_S)$ with 
a probability $P(N_A, N_R,~N_S;t)$,
and the chemical reaction changes the state stochastically.

Each molecule reacts to change its state with a certain transition probability.
Let us consider first a process of $R$ production such that the state 
$(N_A,~N_R,~N_S)$ changes to
$(N_A-1,~N_R+1,~N_S)$.
Since the change is induced by the spontaneous reaction of one $A$ molecule 
among $N_A$ of them to $R$,  
the transition probability of state per time is given by $k_0N_A$.
If the $A$ molecule encounters one $R$, the linear autocatalysis increases
the reaction rate by $k_1$. 
When there are $N_R$ number of $R$ molecules in a well homogenized volume $V$, 
the probability of the encounter is 
equal to the concentration 
$r=N_R/V$. Therefore, the increment
of the transition probability is $k_1 (N_R/V) \times N_A$. 
With a similar consideration, the quadratic autocatalysis
increases the transition probability by $k_2 (N_R/V)^2 \times N_A$.
Thus, the total transition probability of creating one $R$ molecule
is denoted by
\begin{align}
W(N_A,N_R,N_S \rightarrow N_A-1,N_R+1,N_S)&=(k_0+k_1N_R/V+k_2N_R^2/V^2)N_A
\nonumber \\
&=(k_0+\kappa_1N_R+\kappa_2N_R^2)N_A
\label{eq06}
\end{align}
where we define stochastic rate coefficients \cite{lente04,lente05} as
\begin{align}
\kappa_1=k_1/V=k_1c/N, \qquad \kappa_2=k_2/V^2=k_2c^2/N^2
\label{eq07}
\end{align}
with $c=N/V$ being the total concentration of chemically relevant
molecules.
For the recycling process (4) with the rate constant $\lambda$, the transition 
probability is written as
\begin{align}
W(N_A,N_R,N_S \rightarrow N_A+1,N_R-1,N_S)=\lambda N_R .
\label{eq08}
\end{align}
For the $S$ enantiomer, similar transition probabilities are defined.
The time evolution of the probability distribution 
is then written 
by the master equation as
\begin{align}
\frac{d P(N_A,N_R,N_S ;t)}{dt}=
&\sum_{N_R',N_S'} P(N_A',N_R',N_S' ;t)
W(N_A',N_R',N_S' \rightarrow N_A,N_R,N_S)
\nonumber \\
&-\sum_{N_R',N_S'} P(N_A,N_R,N_S ;t)
W(N_A,N_R,N_S \rightarrow N_A',N_R',N_S')
\label{eq09}
\end{align}
where the state $(N_A',~N_R',~N_S')$ is connected to the state 
$(N_A,~N_R,~N_S)$ by the transition probability $W$, and differs the latter 
by numbers of molecules at most by one.
Average concentration of $R$ enantiomer is defined 
by using the probability distribution as 
\begin{align}
\langle r(t) \rangle =\langle N_R(t) \rangle /V 
= V^{-1}\sum_{N_R, N_S} N_R P(N_A,N_R,N_S;t)
\label{eq09a}
\end{align} 
and
its dynamics is described as
\begin{align}
\frac{d\langle r(t) \rangle }{dt}= \langle (k_0+k_1r+k_2r^2)a \rangle -
\lambda \langle r(t) \rangle .
\label{eq10}
\end{align}
Therefore, by neglecting the correlation as 
$\langle ra \rangle \cong \langle r \rangle \langle a \rangle $ and 
$\langle r^2a \rangle \cong \langle r \rangle ^2 \langle a \rangle $,
one recovers the usual rate equations (\ref{eq05}). 

\section{Final probability}

We know that with the recycling process (\ref{eq04}), 
the rate equations (\ref{eq05}) lead to a unique final state.
\cite{saito+04a}
Therefore, one may expect the existence of 
a definite probability distribution $P_f$ associated with
this unique final state.
This $P_f$ can be obtained by numerically simulating the time evolution
(\ref{eq09}), but one can obtain it analytically
by assuming a detailed balance condition
such that the production flow from the state
$(N_A+1,~N_R-1,~N_S)$ to $(N_A,~N_R,~N_S)$ written as
$P_f(N_A+1,N_R-1,N_S)W(N_A+1,N_R-1,N_S \rightarrow N_A,N_R,N_S)$
balances with the counter recycling flow
$P_f(N_A,N_R,N_S)W(N_A,N_R,N_S \rightarrow N_A+1,N_R-1,N_S)$
for $N_R \ge 1$.
This leads to the final probability as
\begin{align}
P_f(N_A,N_R,N_S)
&=\frac{W(N_A+1,N_R-1,N_S \rightarrow N_A,N_R,N_S)}{
W(N_A,N_R,N_S \rightarrow N_A+1,N_R-1,N_S)}P_f(N_A+1,N_R-1,N_S)
\nonumber \\
&=\frac{[k_0+\kappa_1 (N_R-1)+\kappa_2 (N_R-1)^2](N_A+1)}{
\lambda N_R}P_f(N_A+1,N_R-1,N_S)
\nonumber \\
&=\frac{(N_A+N_R)! }{N_R! N_A!}
\frac{ \dis{
\prod_{m=0}^{N_R-1} (k_0+\kappa_1 m+\kappa_2 m^2)}
}{
\lambda^{N_R}}P_f(N_A+N_R,0,N_S).
\label{eq11}
\end{align}
When $N_S=0$, Eq.(\ref{eq11}) determines $P_f(N_A,N_R,0)$ in terms of $P_f(N,0,0)$.
When $N_s \ge 1$, one applies
 similar procedures for $N_S$ again and obtains the relation
\begin{align}
P_f(N_A, N_R,N_S)
=\frac{N!}{N_R! N_S! N_A!}
\frac{\dis{
\prod_{m=0}^{N_R-1} (k_0+\kappa_1 m+\kappa_2 m^2) \prod_{n=0}^{N_S-1} (k_0+\kappa_1 n+\kappa_2 n^2)}
}{
\lambda^{N_R+N_S}}P_f(N,0,0).
\label{eq12}
\end{align}
Here the number conservation $N_A+N_R+N_S=N$ is used.
$P_f(N_A,0,N_S)$ can be obtained from $P_f(N_A,N_R,0)$ by
simply replacing $N_R$ by $N_S$.
The final state  probability distribution is in fact symmetric 
for any $N_R$ and $N_S$ as
\begin{align}
P_f(N_A, N_R,N_S)
=P_f(N_A, N_S,N_R)
\label{eq13}
\end{align}
reflecting the chiral symmetry of the dynamic processes (\ref{eq01})-(\ref{eq04}).
$P_f(N,0,0)$ is determined from the normalization condition
\begin{align}
\sum_{N_R=0}^{N} \sum_{N_S=0}^{N-N_R} P_f(N_A, N_R,N_S)
=\sum_{N_R,N_S} P_f(N-N_R-N_S, N_R,N_S)
=1 .
\label{eq14}
\end{align}
All the reactions except the recycling (\ref{eq04}) consume achiral molecules $A$.
Thus without recycling, reactions proceed and come to a halt 
when all $A$ molecules are
consumed, namely $N_A=0$. The final distribution $P_f$ should hence be zero
for all states with $N_A \ne 0$ as $\lambda \rightarrow 0$.
It is possible by assuming $P_f(N,0,0) \propto \lambda ^N$.
Thus the final probability is written as
\begin{align}
P_f(N_A, N_R,N_S)
={\cal N} \frac{N!}{N_R! N_S! N_A!} \lambda^{N_A} f(k_0,\kappa_1,\kappa_2;N_R) f(k_0,\kappa_1,\kappa_2;N_S)
\label{eq15}
\end{align}
with ${\cal N}$ being the normalization constant, and 
a function $f$ is defined as 
\begin{align}
f(k_0,\kappa_1,\kappa_2;M) =
\begin{cases}
1 & \mbox{for $M=0$},\\
\dis{
\prod_{m=0}^{M-1} (k_0+\kappa_1 m+\kappa_2 m^2)}
 & \mbox{for $M \ge 1$}.
\end{cases}
\label{eq16}
\end{align}
Its dependence on the stochastic rate coefficients $k_0,\kappa_1$ and $\kappa_2$
is explicitly indicated for later convenience. 

For a large system, the probability is denoted as $P_f=e^{-V g(a,r,s)}$
with $r=N_R/V,~s=N_S/V$ and $a=N_A/V=c-r-s$, and
the effective "potential" $g$ is written as
\begin{align}
g(a,r,s)=&-V^{-1} \ln P_f(N_A, N_R,N_S)
\nonumber \\
=&-\mbox{const}+r \ln r +s \ln s + a \ln a -a \ln \lambda 
\nonumber \\
&- \int_0^r dx \ln (k_0+k_1 x+ k_2 x^2)
- \int_0^s dy \ln (k_0+k_1 y+ k_2 y^2).
\label{eq17}
\end{align}
The stationary conditions $\partial g/\partial r=\partial g/\partial s=0$ 
yield 
\begin{align}
(k_0+k_1r+k_2r^2) a = \lambda r, \qquad 
(k_0+k_1s+k_2s^2) a = \lambda s,
\label{eq18}
\end{align}
which are identical with the equations to determine fixed points 
of the rate equations (\ref{eq05}). 
There are racemic solutions $r=s$ which satisfy
\begin{align}
(k_0+k_1r+k_2r^2)(c-2r) = \lambda r
\label{eq19}
\end{align}
and chiral solutions $r \ne s$ for positive $k_2$ as
\begin{align}
r=\frac{c}{2} \Big(X_{\pm}\pm\sqrt{X_{\pm}^2-4\frac{k_0}{k_2c^2}} \Big)
=X_{\pm}-s
\label{eq20}
\end{align}
with
\begin{align}
X_{\pm}=\frac{k_2c^2-k_1c\pm \sqrt{(k_2c^2+k_1c)^2-4k_2c^2 \lambda}}
{2k_2c^2}.
\label{eq21}
\end{align}
Chiral state is impossible to exist
unless quadratic autocatalysis is active.
It is possible only when the obtained values of $r$ and $s$ are real 
and positive.

The probability around the fixed point is characterized
 by the second derivatives;
\begin{align}
g_{rr}=&\frac{\partial^2 g(r,s)}{\partial r^2}
=\frac{1}{r}+\frac{1}{a}-\frac{k_1+2k_2r}{k_0+k_1r+k_2r^2}, \quad 
g_{rs}=\frac{\partial^2 g(r,s)}{\partial r \partial s}=\frac{1}{a},
\nonumber \\
g_{ss}=&\frac{\partial^2 g(r,s)}{\partial s^2}
=\frac{1}{s}+\frac{1}{a}-\frac{k_1+2k_2s}{k_0+k_1s+k_2s^2}.
\label{eq22}
\end{align}
Especially around the racemic fixed point, which we denote here 
$r^*=s^*$, symmetry implies $g_{rr}=g_{ss}$,
and the distribution of fluctuations $\delta r=r-r^*$ and  $\delta s=s-s^*$
is approximately given as
\begin{align}
P_f(r,s)&\approx e^{-Vg(r,s)}
\nonumber \\
&=\exp \Big\{-V\Big[ \frac{1}{4}(g_{rr}+g_{rs})(\delta r + \delta s)^2
+\frac{1}{4}(g_{rr}-g_{rs})(\delta r - \delta s)^2 \Big] \Big\}.
\label{eq23}
\end{align}
Therefore, the fluctuations are determined with $a^*=c-r^*-s^*$ as
\begin{align}
\langle (\delta r+\delta s)^2 \rangle &= \frac{2}{V(g_{rr}+g_{rs})}
= \frac{2a^*r^*(k_0+k_1r^*+k_2r^{*2})}{V(k_0c+(c-a^*)k_1r^*+(c-2a^*)k_2r^{*2})},
\nonumber \\
\langle (\delta r-\delta s)^2\rangle &= \frac{2}{V(g_{rr}-g_{rs})}
= \frac{2r^*(k_0+k_1r^*+k_2r^{*2})}{V(k_0-k_2r^{*2})}.
\label{eq24}
\end{align}
These formulae are useful in the following detailed studies of
the final probability distribution for some typical cases.

\subsection{Spontaneous production with and without recycling}
Only with a spontaneous production and a recycling ($k_0, \lambda >0$ and
$k_1=k_2=0$), the final probability distribution is easily calculated as
\begin{align}
P_f(N_A, N_R,N_S)
=\frac{N!}{N_R! N_S! N_A!} \frac{k_0^{N_R+N_S}\lambda^{N_A}}{(2k_0+\lambda)^N}
\label{eq25}
\end{align}
with $N=N_A+N_R+N_S$.
It has a peak at a racemic point obtained from Eq.(\ref{eq19}) as
\begin{align}
r^*=s^*=\frac{k_0}{2k_0+\lambda}c, \qquad 
a^*=c-r^*-s^*=\frac{\lambda}{2k_0+\lambda}c
\label{eq26}
\end{align}
and the probability has a Gaussian profile in the vicinity of the peak as
\begin{align}
P_f(N_A, N_R,N_S) \approx \exp \Big\{-V \Big[
\frac{(2k_0+\lambda)^2}{4k_0 \lambda c} (\delta r+\delta s)^2
+\frac{2k_0+\lambda}{4k_0 c} (\delta r-\delta s)^2\Big] \Big\}
\label{eq27}
\end{align}
with $\delta r=r-r^*$ and $\delta s=s-s^*$.
The peak position $(r^*,s^*)$ agrees with the fixed point 
of the rate equations (\ref{eq05}).
Fluctuations $\delta r$ and $\delta s$ are anisotropic such that 
in the direction $r=-s$ fluctuation vanishes as $\lambda \rightarrow 0$ whereas in the direction $r=s$ finite fluctuation remains. 

In the limit of vanishing recycling ($\lambda \rightarrow 0$), 
there is no reason that the final probability distribution takes the
form given by the detailed balance condition, but 
it is interesting to examine $P_f$ in this limit. 
Since all the substrate molecules turn into chiral products, 
$N_A$ should be zero 
so that  $P_f(N_A >0, N_R, N_S)=0$.
Thus, the final state lies on a fixed line $r+s=c$ or $N_R+N_S=N$.
The probability (\ref{eq25}) in this limit takes a binomial distribution 
along the fixed line as 
\begin{align}
P_f(0, N_R,N-N_R)
=\frac{N!}{N_R! (N-N_R)! } \Big(\frac{1}{2} \Big)^N.
\label{eq28}
\end{align}
This result is identical with the solution obtained by Lent, 
who analyzed the same system with no recyling assuming a complete achiral state ($N,0,0)$ as the initial condition
\cite{lente04,lente05}.

\subsection{Linear autocatalysis with and without recycling}
Addition of a linear autocatalytic process does not alter the final probability
distribution qualitatively.
With $k_2=0$, the probability has a peak at a racemic point obtained from 
Eq.(\ref{eq19}) as
\begin{align}
r^*=s^*=c
\frac{\sqrt{(2k_0-k_1c+\lambda)^2+8k_0k_1c}-(2k_0-k_1c+\lambda)}{4k_1c}
\label{eq29}
\end{align}
which corresponds to a fixed point of the rate equations (\ref{eq05}).
Fluctuations around the fixed point given by Eq.(\ref{eq24}) are non-negative
as long as $k_2=0$.

If there is no recycling ($\lambda=0$), a fixed point is at 
$r^*=s^*=c/2$ and $a^*=0$.
Thus the fluctuation $ \langle (\delta r+\delta s)^2 \rangle$ vanishes
and $r$ and $s$ fluctuate along the diagonal line $r+s=c$.
In this case, fluctuations 
$ \langle (\delta r)^2 \rangle = \langle (\delta s)^2 \rangle$
are still finite as long as $k_0 >0$.
Only when the spontaneous production is absent ($k_0=0$),
fluctuations diverge, indicating that a distribution becomes flat
along  a ridge $r+s=c$.

In fact, for $\lambda \rightarrow 0$, Eq.(\ref{eq15}) tells us that
the final probability is nonzero only along the fixed line 
$N_A = 0$, or $N_R+N_S=N$, in the $N_R$-$N_S$ phase space,
and the normalized probability is calculated explicitly as
\begin{align}
P_f(0, N_R,N-N_R)
=\frac{N!}{N_R! (N-N_R)! } 
\frac{f(k_0,\kappa_1,0;N_R)f(k_0,\kappa_1,0;N-N_R)}
{f(2k_0,\kappa_1,0;N)}.
\label{eq30}
\end{align}
This result is again identical with the solution obtained by Lent, 
starting from the complete achiral state ($N,0,0)$\cite{lente04,lente05}.

If $k_0$ is as small as $k_0=\kappa_1=k_1/V$, 
then $f(k_0,k_0,0;M)=k_0^MM!$ and $f(2k_0,k_0,0;N)=k_0^N(N+1)!$,
 and the final probability
distribution is constant as $P_f(0,N_R,N-N_R)=1/(N+1)$.
For further smaller $k_0$, 
$P_f$ has double peaks of height $P_f\cong 1/2$ at $N_R=0$ or
$N$, and essentially vanishes otherwise. 
The peaks at $N_R=0$ or $N_S=0$ mean that an enantiomer first produced
by the spontaneous reaction converts all the achiral substrate molecules into
her type of chirality by linear autocatalysis before the second type of enantiomer
is produced spontaneously; this is just a single mother scenario.
The former process takes place within a time about $\sum_{N_R=1}^{N-1} 1/\kappa_1 N_R (N-N_R) \sim 2 \ln N/ \kappa_1 N$ whereas
the latter requires the time $1/k_0N$. 
Therefore, by neglecting the logarithmic correction,
the second mother has no chance to be born 
when the spontaneous production
rate $k_0$ is smaller than $\kappa_1$, 

\subsection{Quadratic autocatalysis with and without recycling}

When $k_2 > 0$, there are at most three racemic fixed points.
Those points with $r^* > \sqrt{k_0/k_2}$ are unstable, since
the fluctuation $\langle (\delta r-\delta s)^2\rangle$ 
given in Eq.(\ref{eq24}) becomes negative.
The probability distribution is not maximum there but represents
a saddle point or a valley structure.
On the other hand, 
new chiral fixed points bifurcate from $r=s=\sqrt{k_0/k_2}$
 when $X_{\pm}^2 \ge 4k_0/k_2c^2$,
 as is described in Eq.(\ref{eq20}).
There are at most four chiral fixed points,
depending on the rate coefficients.
For $k_1=0$, the critical value of the quadratic rate coefficient $k_{2c}$
is calculated as
\begin{align}
k_{2c}c^2= 4k_0(1+\frac{\lambda}{4k_0})^2 .
\label{eq31}
\end{align}
When $k_2$ is larger than $k_{2c}$, the racemic fixed point is unstable,
and correspondingly the final distribution $P_f$ has double peaks
associated with chiral states.

When $\lambda \rightarrow 0$, 
the final probability distribution is expected to vanish for $N_A > 0$.
On the line $N_A=0$, it is given by $P_f(0,N_R,N-N_R)$ in Eq.(\ref{eq15})
but we are unsuccessful so far to obtain a closed analytic expression 
of the normalization factor ${\cal N}$.
When $k_1=\lambda=0$, $P_f$ has peaks at 
chiral fixed points $(r_+,s_-=c-r_+)$ and ($r_-,s_+=c-r_-)$ with
\begin{align}
r_{\pm}=c\frac{1 \pm \sqrt{1-4k_0/k_2c^2}}{2}=s_{\mp}.
\label{eq32}
\end{align}
The ee order parameter takes the value
\begin{align}
\phi=\frac{r_{\pm}-s_{\mp}}{r_{\pm}+s_{\mp}}
= \pm \sqrt{1-\frac{4k_0}{k_2c^2}}.
\label{eq33}
\end{align}

\section{Numerical Integration of the Master Equation}

In this section we discuss on the time evolution of the probability distribution.
It is obtained by integrating
the master equation (\ref{eq09}) by means of the fourth-order
 Runge-Kutta method
 \cite{press+92}.
Similarly to the previous section, three typical cases are
discussed in the following subsections.
For the numerical integration, we consider
a system with the total number of active chemical species to be $N=100$.

\begin{figure}[h]
\begin{center} 
\includegraphics[width=0.44\linewidth]{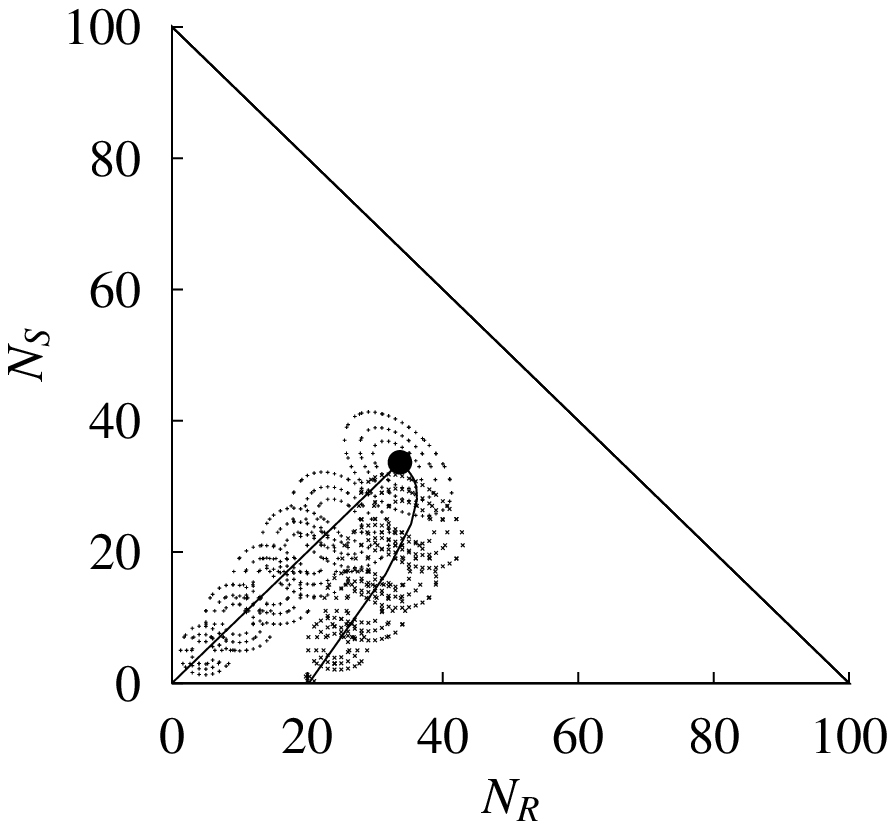}
\includegraphics[width=0.55\linewidth]{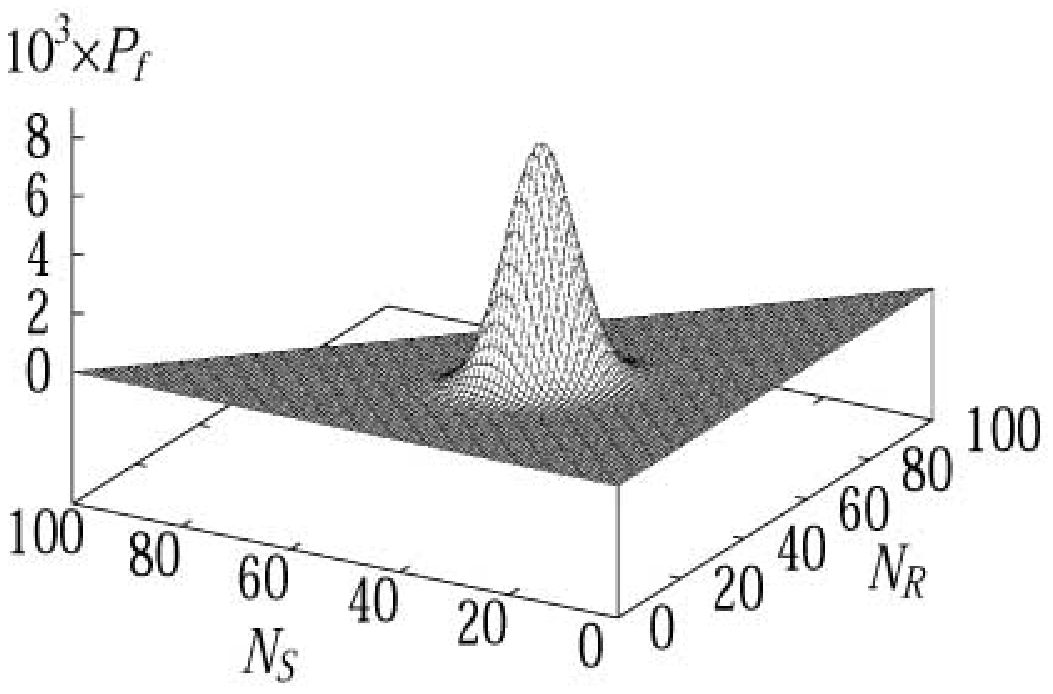}\\
(a) \hspace{8cm} (b) \\
\includegraphics[width=0.44\linewidth]{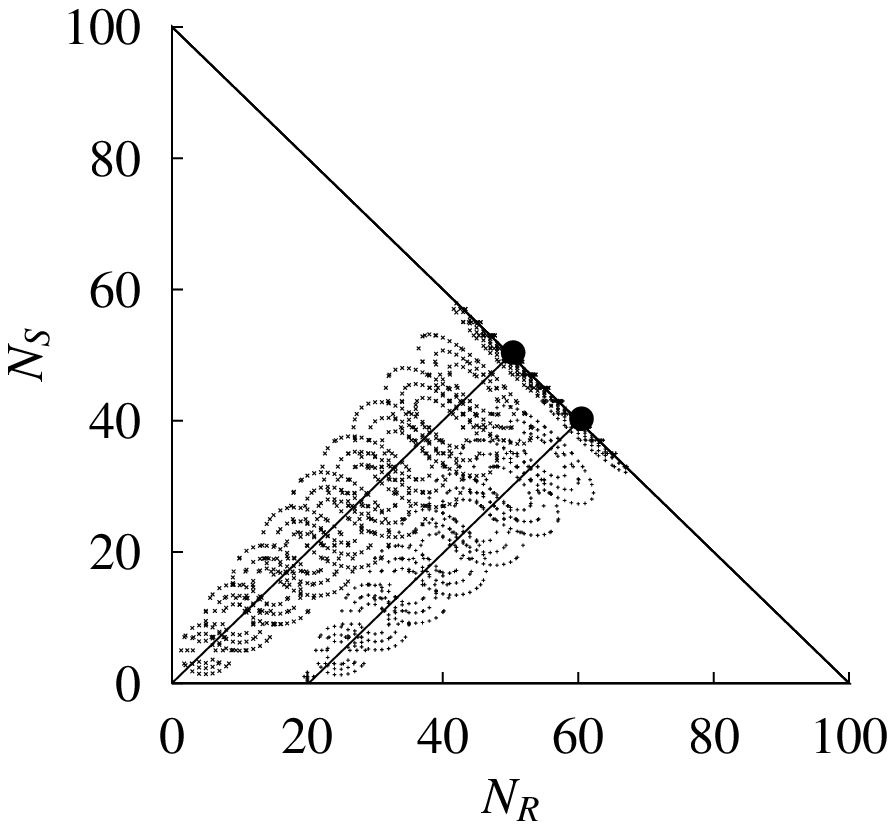}
\includegraphics[width=0.55\linewidth]{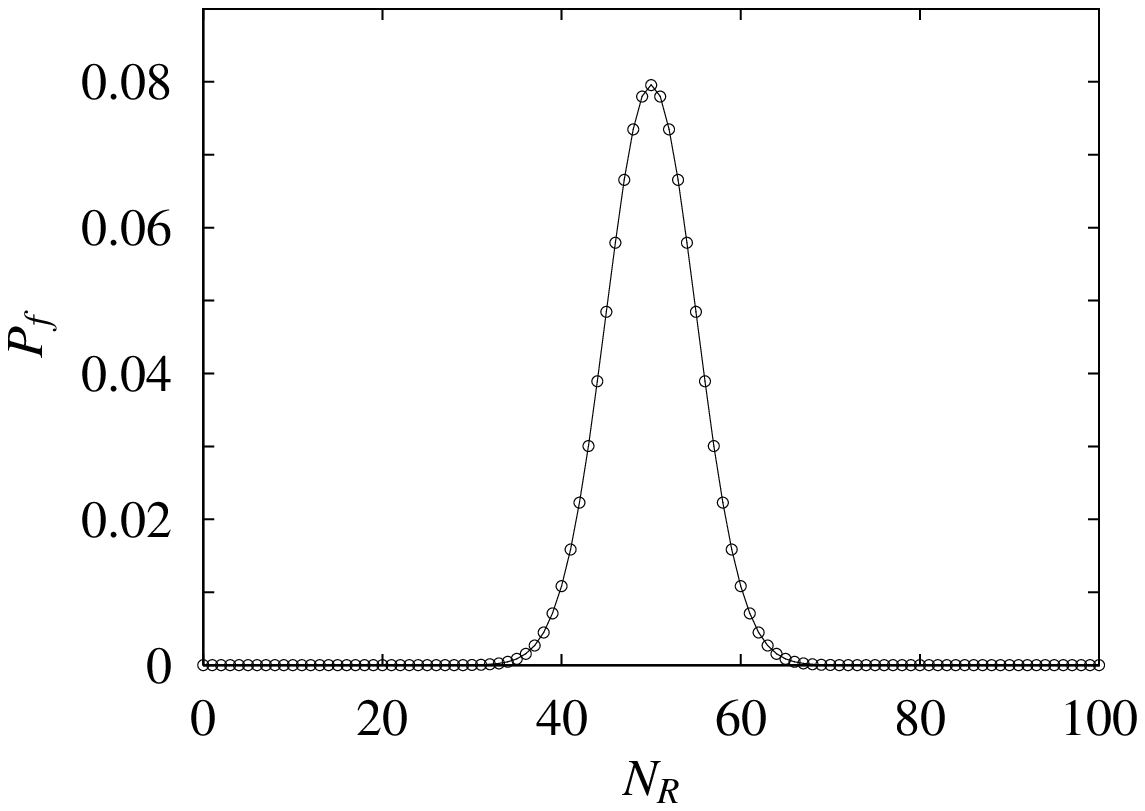}\\
(c) \hspace{8cm} (d) 
\end{center} 
\caption{Trajectories of probability contours obtained by 
numerically integrating the master equation (\ref{eq09}) 
for spontaneous production
(a) with ($\lambda=k_0$) and (c) without ($\lambda=0$) recycling.
(b) and (d) represent final probability distributions with and without
recycling, respectively. 
Other parameters are  $k_1=k_2=0$ with $N=100$.
}
\label{fig1}
\end{figure}

\subsection{Spontaneous production}

With a spontaneous production $k_0$  and a finite recycling as $\lambda=k_0$
(and $k_1=k_2=0$), 
time evolution of the probability distribution is depicted in Fig.1(a)
by contour lines
of a relative height interval with a quarter of the maximum height
at several times.
Whether the system starts from a complete achiral state
$(N_A,N_R,N_S)=(100,0,0)$ or a chiral state $(80,20,0)$, 
probability distributions keep a single-peak profile and  
converge to the unique final distribution.
The final probability distribution shown 
in Fig.1(b) agrees with the result given by Eq.(\ref{eq25}).
The peak position of the transient probability distribution 
follows the trajectory determined by the rate equations,
drawn by continuous curves in Fig.1(a).
They approach a racemic fixed point 
$N_R^*=N_S^*=N/3$
 for $k_0=\lambda$.

Without recycling ($\lambda=0$), achiral substrate is consumed up ultimately, 
and the final state has a finite probability only at $N_A=0$ or on a line
$N_R+N_S=N=100$.
Also the remarkable fact is that without recycling
the final state depends on the initial condition, as shown in Fig.1(c);
the probability distribution started from an initially achiral state
$N_R(0)=N_S(0)=0$ keeps its center on a racemic state $N_R=N_S$
during the evolution, 
whereas that from a chiral initial state remains chiral
even though the ee order parameter decreases.
The final probability distribution given in the binomial form Eq.(\ref{eq28})
is valid only for a racemic state, as shown in Fig.1(d).

\begin{figure}[h]
\begin{center} 
\includegraphics[width=0.44\linewidth]{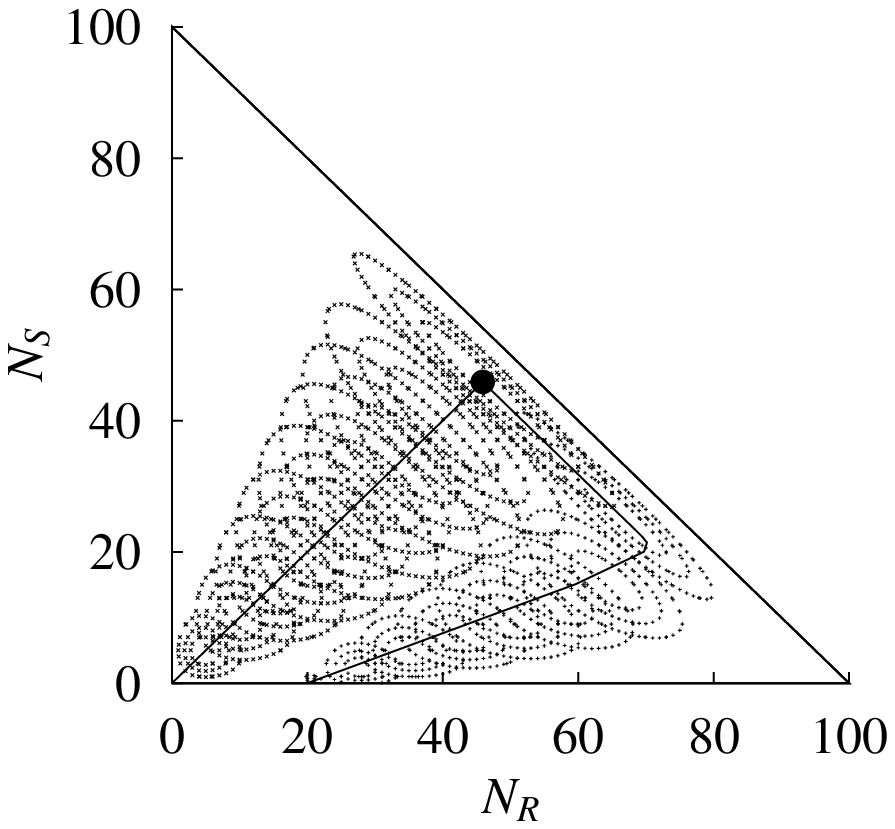}
\includegraphics[width=0.55\linewidth]{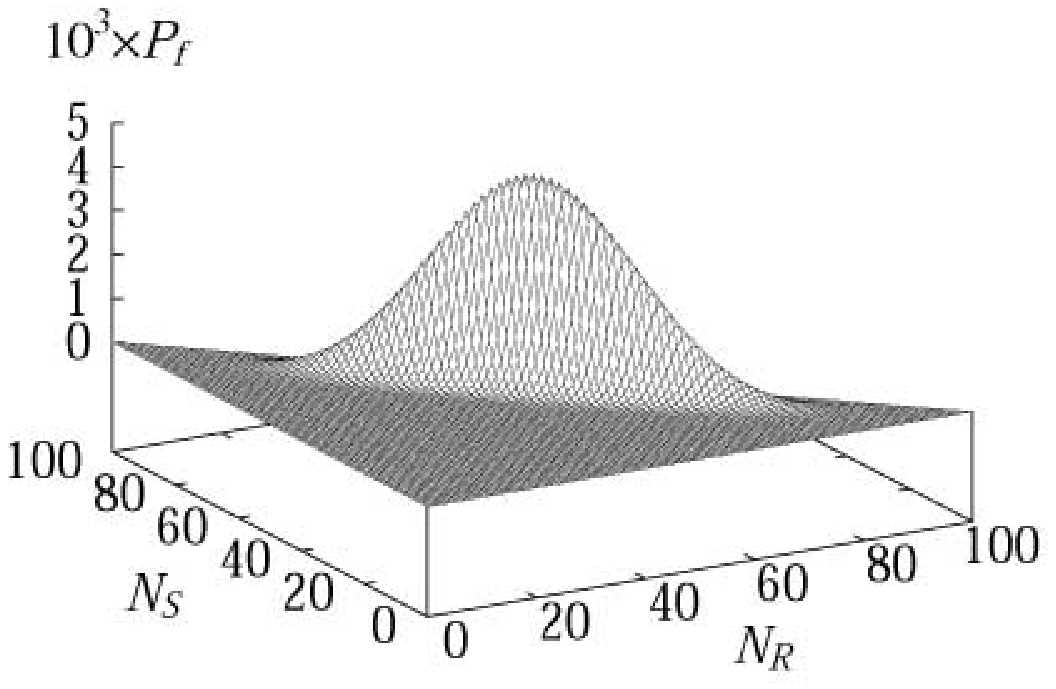}\\
(a) \hspace{8cm} (b) \\
\includegraphics[width=0.44\linewidth]{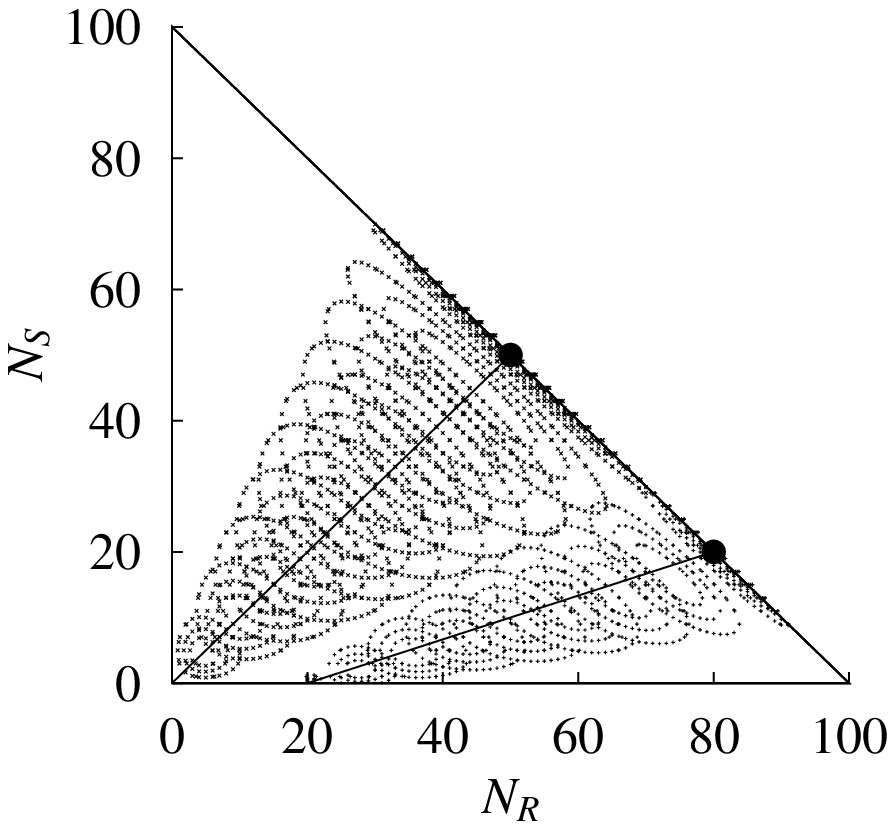}
\includegraphics[width=0.55\linewidth]{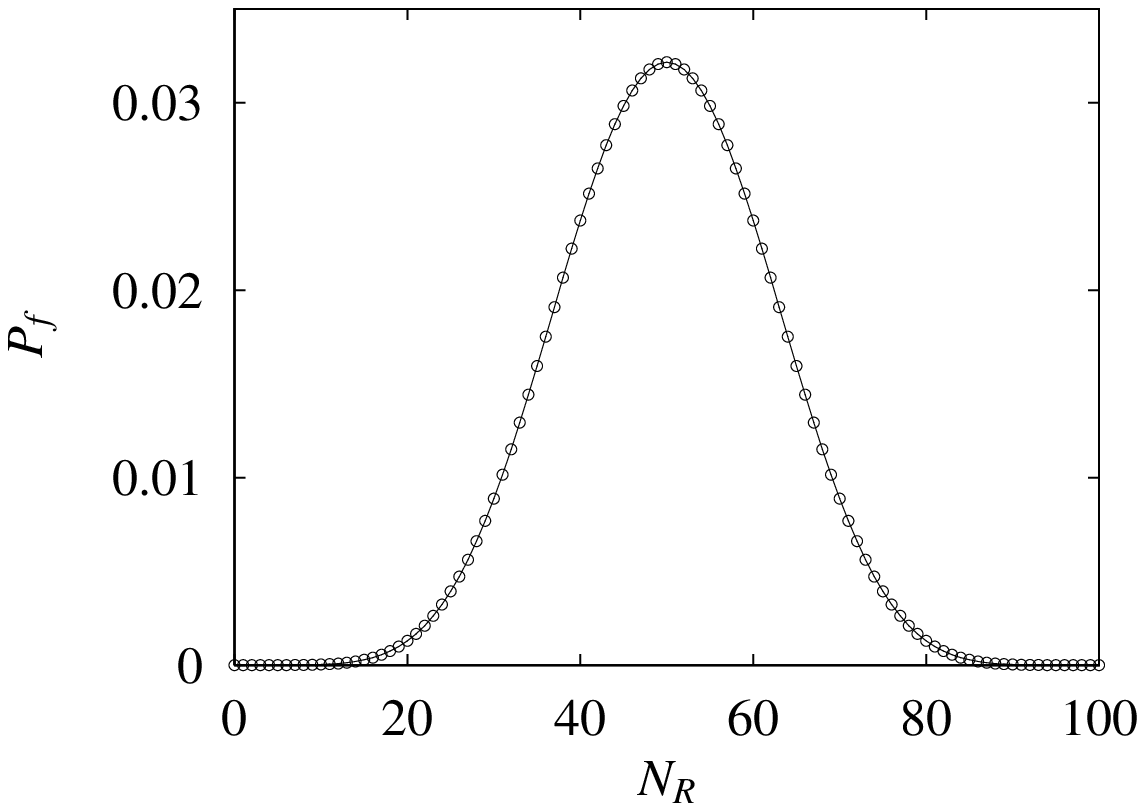}\\
(c) \hspace{8cm} (d) 
\end{center} 
\caption{Trajectory of the probability contour for 
linear autocatalyses ($\kappa_1=10^{-1}k_0$ )
(a) with ($\lambda=k_0$ )
and (c) without ($\lambda=0$ ) recycling.
(b) and (d) represent final probability distributions with and without
recycling, respectively. 
Other parameters are  $k_2=0$ with $N=100$.
}
\label{fig2}
\end{figure}

\subsection{Linear autocatalysis}
We now include a linear autocatalytic process with $\kappa_1=0.1k_0$.
With a finite recycling as $\lambda=k_0$, the probability distribution converges
to a unique profile with a single peak at a racemic state, as shown in Fig.2(a): 
 at the initial stage
 the system started from a chiral state with $N_R(0)=20$ and $N_S(0)=0$ 
behaves differently from that
started from an achiral state, but ultimately it turns to converge to
a stable racemic fixed point.
The final probability distribution is the one given in Eq.(\ref{eq15}), 
as shown in Fig.2(b).

When there is no recycling, 
the final probability depends on the initial condition, as shown in Fig.2(c).
The final probability staring from the achiral state $N_R=N_S=0$
is well described by the profile Eq.(\ref{eq30}) 
determined by the detailed balance condition, as shown in Fig.2(d).

\begin{figure}[h]
\begin{center} 
\includegraphics[width=0.32\linewidth]{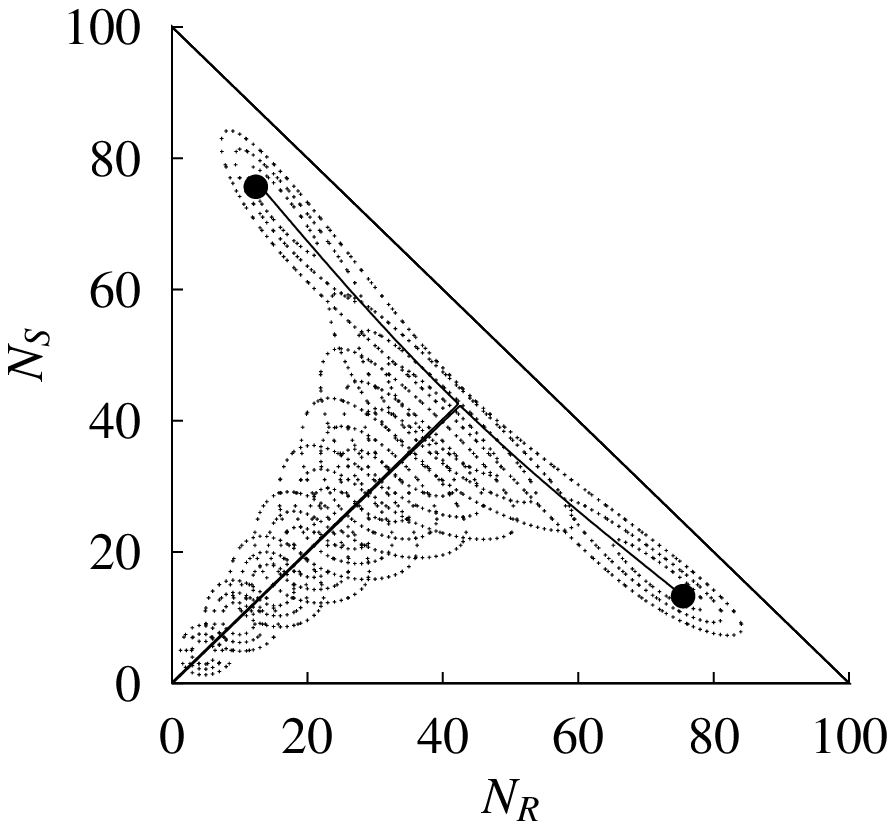}
\includegraphics[width=0.43\linewidth]{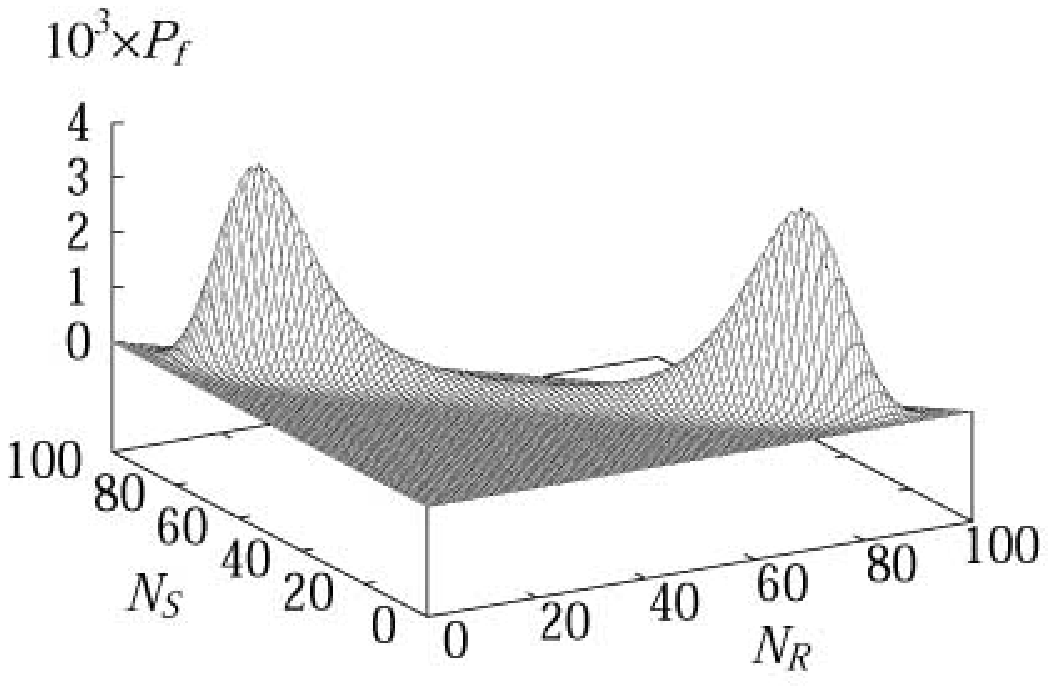}\\
(a) \hspace{8cm} (b) \\
\includegraphics[width=0.32\linewidth]{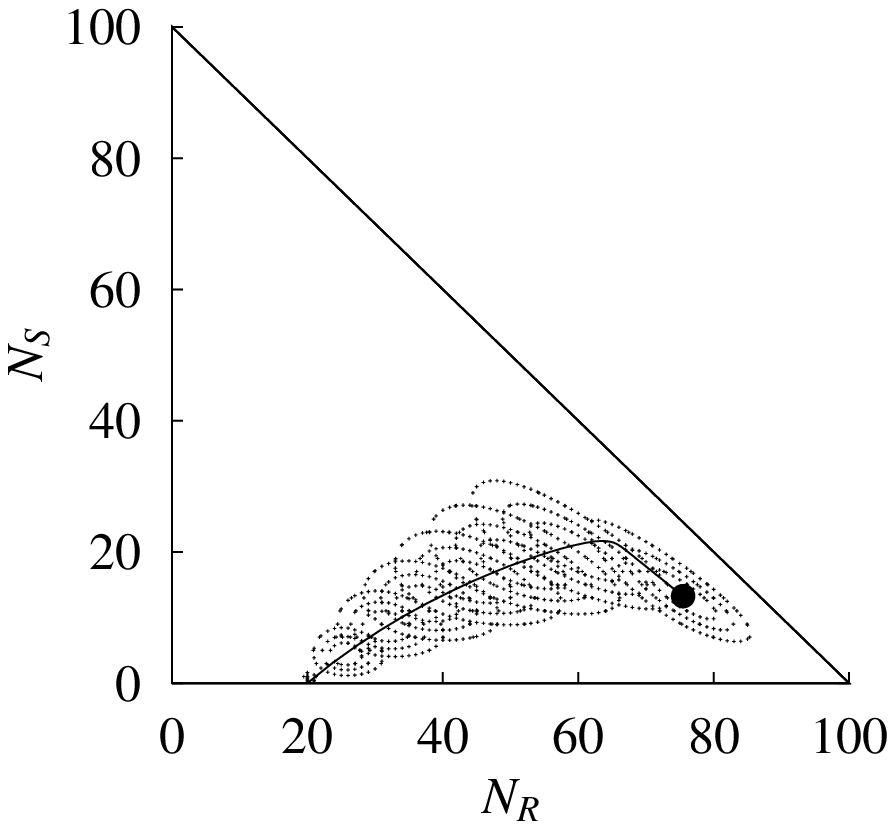}
\includegraphics[width=0.43\linewidth]{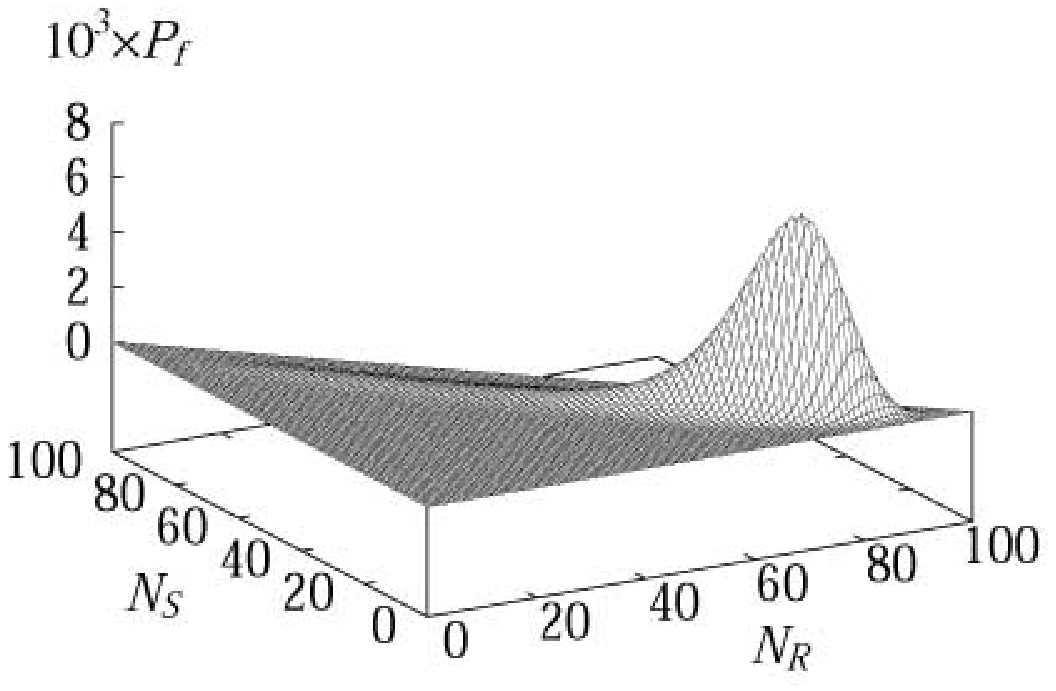}\\
(c) \hspace{8cm} (d) \\
\includegraphics[width=0.32\linewidth]{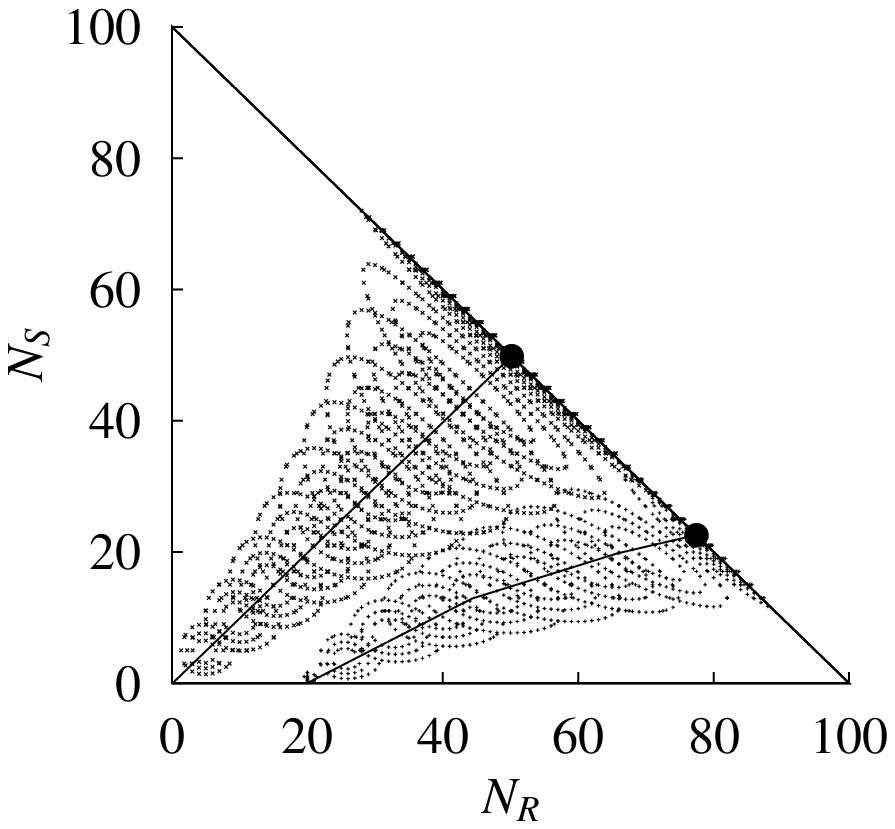}
\includegraphics[width=0.41\linewidth]{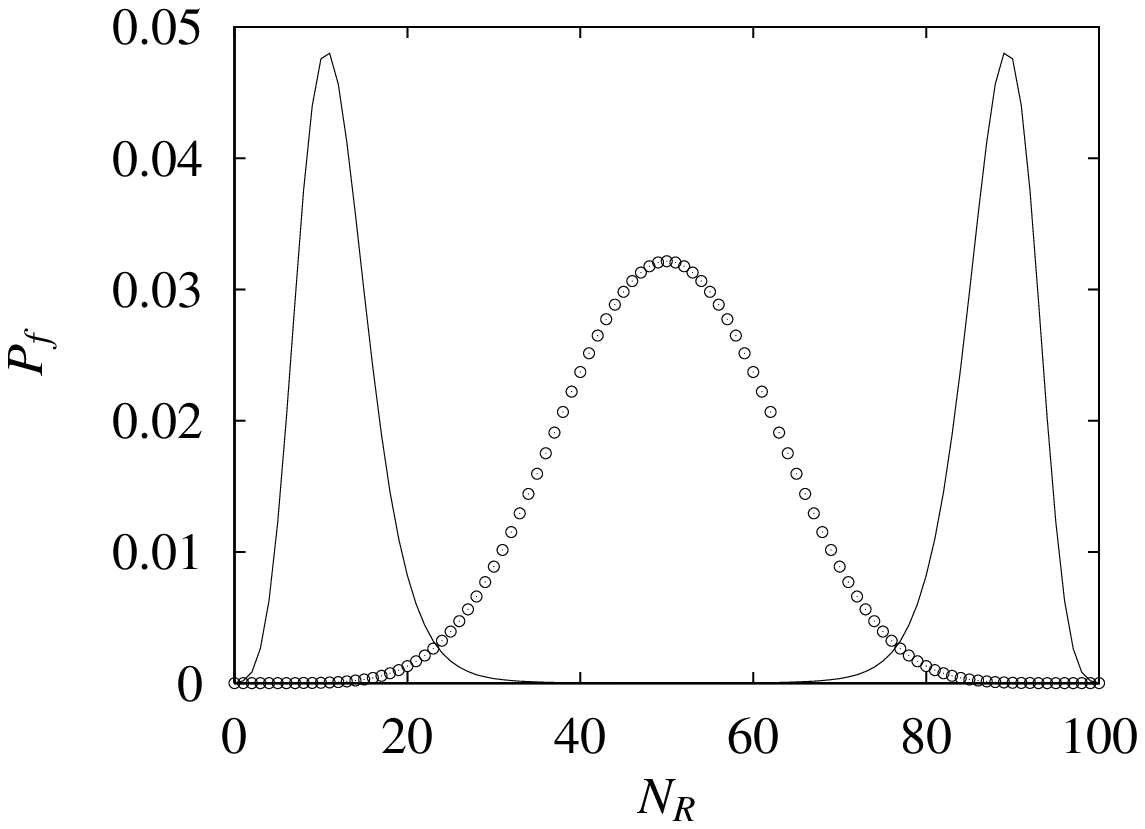}\\
(e) \hspace{8cm} (f) 
\end{center} 
\caption{Trajectory of the probability contour for 
nonlinear autocatalysis ($\kappa_2=10^{-3}k_0$)
(a), (c) with ($\lambda=k_0$) and (e) without ($\lambda=0$) recycling.
Initial state for (a) is achiral, and for (c) is chiral.
(b),(d) and (f) represent the final probability distributions 
corresponding to (a), (c) and (e), respectively.
In (e) the final distribution obtained by numerical integration, represented
by symbols, is quite different from the expected curve from the detailed balance 
condition, Eq.(\ref{eq15}). 
Other parameters are  $k_1=0$ with $N=100$.
}
\label{fig3}
\end{figure}

\subsection{Quadratic autocatalysis}

With a quadratic autocatalysis, the probability distribution
behaves qualitatively
differently from the previous cases.
Even with a recycling process, the probability profile depends on the initial
condition in the sense explained below.
If the system starts from a completely achiral state without
chiral species, $N_R=N_S=0$, the probability distribution
is symmetric due to the symmetric dynamics, 
but the initial single peak splits around the unstable racemic fixed
point, and the double-peak structure develops slowly, 
as shown in the contour evolution in Fig.3(a).
The parameters chosen are  $\kappa_2=10^{-3}k_0, \kappa_1=0$ and $\lambda=k_0$.
The final probability has a double-peak as shown in Fig.3(b), 
and the profile is described by Eq.(\ref{eq15}).
On the contrary, when the system starts from a chiral state, 
as shown in Fig.3(c), the probability distribution approaches a single peak 
profile, as shown in Fig.3(d); the peak position
corresponds to one of the double-peak structure in the symmetric case in
Fig.3(a).
The probability distribution accompanies a long tail in the direction 
of the other peak position, but 
the second peak does not develop during the numerical integration upto the time
of order $6/\lambda$. This resembles to the symmetry breaking phase transition
in equilibrium systems, and we may call that the system undergoes a
chiral symmetry breaking.

Without recycling ($\lambda=0$), the evolution of the 
system stops when the whole substrate 
molecules $A$ are consumed  $N_R+N_S=N$, 
as shown in Fig.3(e), quite similar to the previous two cases.
However, the final probability started from an achiral state, 
shown by symbols in Fig.3(f), is completely different from the 
$\lambda \rightarrow 0$ limit of Eq.(\ref{eq15}), 
which is drawn by a continuous curve.
It is not impossible, since at $\lambda=0$ we cannot rely on 
the detailed balance condition between the production and the recycling
 reaction.
Then, what is the final probability distribution?
We shall discuss this problem in the second next section.

\section{Eigenvalue Analysis  of the Master Equation}

Before describing the probability distribution of a system without recycling,
we consider the symmetry breaking for a nonlinear autocatalytic system
with recycling in terms of the degeneracy in eigenvalues of
the master equation.

The master equation (\ref{eq09}) is a linear differential equation for the
probability distribution,
rewritten by using an evolution matrix $\mathbf M$  as
\begin{align}
\frac{d P(N_A,N_R,N_S ;t)}{dt}=
\sum_{N_A',N_R',N_S'} (N_A,N_R,N_S|{\mathbf M}|N_A',N_R',N_S') 
 P(N_A',N_R',N_S' ;t).
 \label{eq34}
\end{align} 
Here the evolution matrix $\mathbf M$  
is related to 
the transition probabilities $W$, 
defined by Eqs.(\ref{eq06}) and (\ref{eq08}),
as
\begin{align}
&(N_A,N_R,N_S|{\mathbf M}|N_A',N_R',N_S') =
\nonumber \\
&
\begin{cases}
W(N_A',N_R',N_S' \rightarrow N_A,N_R,N_S) & \mbox{for $(N_A',N_R',N_S') \ne (N_A,N_R,N_S)$}, \\
- \displaystyle{\sum_{N_{A1}, N_{R1},N_{S1}}} W(N_A,N_R,N_S \rightarrow N_{A1},N_{R1},N_{S1}) 
 & \mbox{for $(N_A',N_R',N_S') = (N_A,N_R,N_S)$} \\
\end{cases}
\label{eq35}
\end{align} 
and the matrix element is nonvanishing between states $(N_A,N_R,N_S)$
and $(N_A',N_R',N_S')$ with differences of numbers of molecules at
most by one and $N_A+N_R+N_S=N_A'+N_R'+N_S'=N$.
By the use of eigenfunctions $\Psi_i$ and eigenvalues $\Lambda_i$
 of the matrix ${\mathbf M}$ as given by 
\begin{align}
{\mathbf M} \Psi_{i}= \Lambda_i \Psi_{i}
\label{eq36}
\end{align} 
the time development of the probability distribution is expressed 
by a series as
\begin{align}
P(N_A,N_R,N_S ;t)= \sum_{i=0}^{\infty} a_{i} e^{\Lambda_i t} \Psi_i.
\label{eq37}
\end{align} 
Since the probability distribution  satisfies the conservation 
of the probability as $\sum_{N_R,N_S}P=1$,
all the $(N+1)(N+2)/2$ eigenvalues $\Lambda_i$ must be non-positive.
Therefore, we order
eigenvalues as $\Lambda_0=0 \ge \Lambda_1 \ge \cdots \Lambda_i \ge 
\Lambda_{i+1} \cdots $.
An equilibrium distribution $P_f$, if exists, is to be given by
the eigenstate $\Psi_0$ of the nondegenerate eigenvalue $\Lambda_0=0$.  
We analyze three typical cases again.

\subsection{Spontaneous production}
In this case, the matrix $\mathbf M$ is simplified by introducing matrices
$\mathbf A$, $\mathbf A^*$, 
$\mathbf R$, $\mathbf R^*$, 
$\mathbf S$, $\mathbf S^*$ which have 
mostly zero elements except
the following nonzero ones as
\begin{align}
&(N_A-1,N_R,N_S|{\mathbf A}|N_A,N_R,N_S) =N_A, \quad
(N_A+1,N_R,N_S|{\mathbf A^*}|N_A,N_R,N_S) =1, 
\nonumber \\
&(N_A,N_R-1,N_S|{\mathbf R}|N_A,N_R,N_S) =N_R, \quad
(N_A,N_R+1,N_S|{\mathbf R^*}|N_A,N_R,N_S) =1, 
\nonumber \\
&(N_A,N_R,N_S-1|{\mathbf S}|N_A,N_R,N_S) =N_S, \quad
(N_A,N_R,N_S+1|{\mathbf S^*}|N_A,N_R,N_S) =1.
\label{eq38}
\end{align}
These matrices satisfy the following nonzero commutation relations 
\begin{align}
[\mathbf A, \mathbf A^*]= [\mathbf R, \mathbf R^*] = [\mathbf S, \mathbf S^*]=1
\label{eq39}
\end{align}
and other commutators vanish. 
Regarding the state (0,0,0) as a vacuum, we obtain
$(\mathbf A^*)^{N_A}|0,0,0)=|N_A,0,0)$,
and thus $\mathbf A^*$ may be regarded as an normalized creation operator 
of $A$ molecules.
Similar interpretation holds for other matrices, and 
thus three bilinear products 
$\mathbf A^* \mathbf A,\, \mathbf R^* \mathbf R,\,
\mathbf S^* \mathbf S$ represent number operators.
The time-evolution matrix $\mathbf M$ is expressed as 
a bilinear form of these matrices as
\begin{align}
\mathbf M=(\mathbf R^*- \mathbf A^*)(k_0 \mathbf A- \lambda \mathbf R)+
(\mathbf S^* - \mathbf A^*)(k_0 \mathbf A- \lambda \mathbf S).
\label{eq40}
\end{align}
By introducing new combinations defined as 
\begin{align}
&\mathbf C_0=\mathbf A +\mathbf R+\mathbf S, \quad 
\mathbf C_0^*=\frac{\lambda \mathbf A^* + k_0 \mathbf R^* + k_0 \mathbf S^*}{2k_0+\lambda},
\nonumber \\
&\mathbf C_1=\mathbf R-\mathbf S, \quad 
\mathbf C_1^*=\frac{\mathbf R^* -  \mathbf S^*}{2},
\nonumber \\
&\mathbf C_2=-2k_0 \mathbf A +\lambda \mathbf R+ \lambda \mathbf S, \quad 
\mathbf C_2^*=\frac{-2 \mathbf A^* + \mathbf R^* + \mathbf S^*}{4k_0+ 2\lambda},
\label{eq41}
\end{align}
one can prove that these new matrices 
satisfy the commutation relations similar to Eq.(\ref{eq39})
as $[\mathbf C_i,\mathbf C^*_j]=\delta_{ij}$,
$[\mathbf C_i,\mathbf C_j]=[\mathbf C^*_i,\mathbf C^*_j]=0$, 
and the time evolution matrix is represented in a diagonal form as
\begin{align}
\mathbf M=- \lambda \mathbf C_1^* \mathbf C_1 -(2k_0+\lambda)\mathbf C_2^* \mathbf C_2.
\label{eq42}
\end{align}
$\mathbf M$'s independence of $\mathbf C_0^* \mathbf C_0$ is related to the
total number conservation in the present system.
Since the eigenvalues of 
$\mathbf C_1^* \mathbf C_1$ and $\mathbf C_2^* \mathbf C_2$ are 
non-negative integers $n_1,~n_2 = 0, 1, 2 \cdots$, and $n_1+n_2 \le N$,
the eigenvalues of the time evolution is written as
\begin{align}
\Lambda_{n_1,n_2}= - \lambda n_1 -(2k_0+\lambda)n_2.
\label{eq43}
\end{align}
The zero eigenvalue $\Lambda_{00}$ is nondegenerate as long as $\lambda>0$,
and the system relaxes to a unique final state:
No phase transition is expected.

On the other hand, without recycling $\lambda=0$ there is a
large degeneracy in the eigenvalues. 
Especially for a zero eigenvalue
there are $N+1$ degeneracy since $\Lambda_{n_1,0}=0$ for 
any $n_1$ from 0 to $N$.
It represents the non-ergodicity of the
system and the final state depends on the initial condition, as is anticipated
from the rate equation analysis.

\subsection{Linear autocatalysis}

With a linear autocatalysis, we have failed so far to obtain
analytic expressions of eigenvalues of the time-evolution
matrix $\mathbf M$, 
and we have to rely on the numerical method.
We use the subroutine "dgeev" in LAPACK 
to obtain eigenvalues and eigenfunctions of a general matrix, 
but because of the algorithmic restriction, the maximum size of $N$
is limited to 50.
Because of the smaller size $N$ in this section than the one in the
previous sections, 
the parameters are chosen to be a little larger as 
$\kappa_1=0.2k_0,~ k_2=0,~\lambda=k_0$.
First few largest eigenvalues
are depicted in Fig.4(a) as a function of 
the total number of chemical reactants $N$.
With a finite recycling $\lambda >0$, the second largest eigenvalue 
$\Lambda_1$ increases with $N$,
 but seems to saturate before it approaches $\Lambda_0=0$.
This finite gap between $\Lambda_1$ and 0 means the absence of non-ergodicity
and of the phase transition.
The 0th eigenfunction Fig.4(b) looks similar to the final probability
distribution for a larger system with $N=100$, shown in Fig.2(b). 
The 1st eigenfunction in Fig.4(c)
is assymmetric around $N_R=N_S$ line.

\begin{figure}[h]
\begin{center} 
\includegraphics[width=0.49\linewidth]{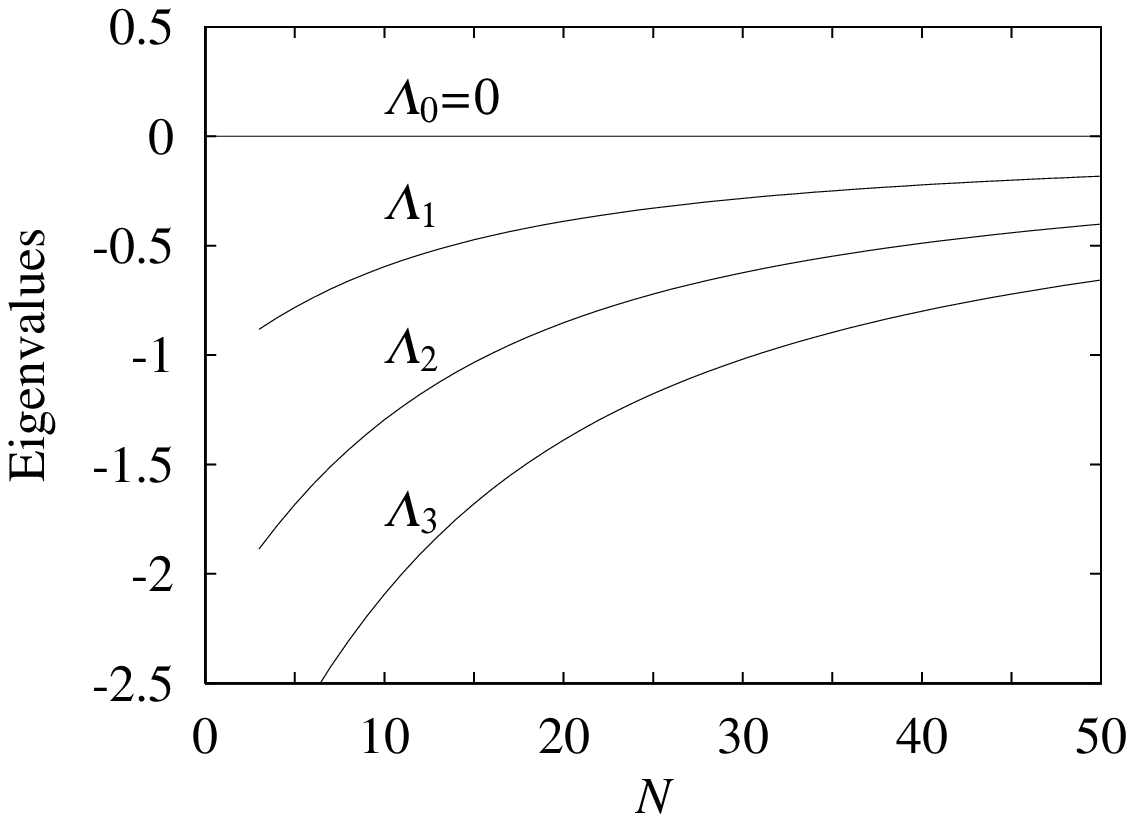}\\
(a)\\
\includegraphics[width=0.49\linewidth]{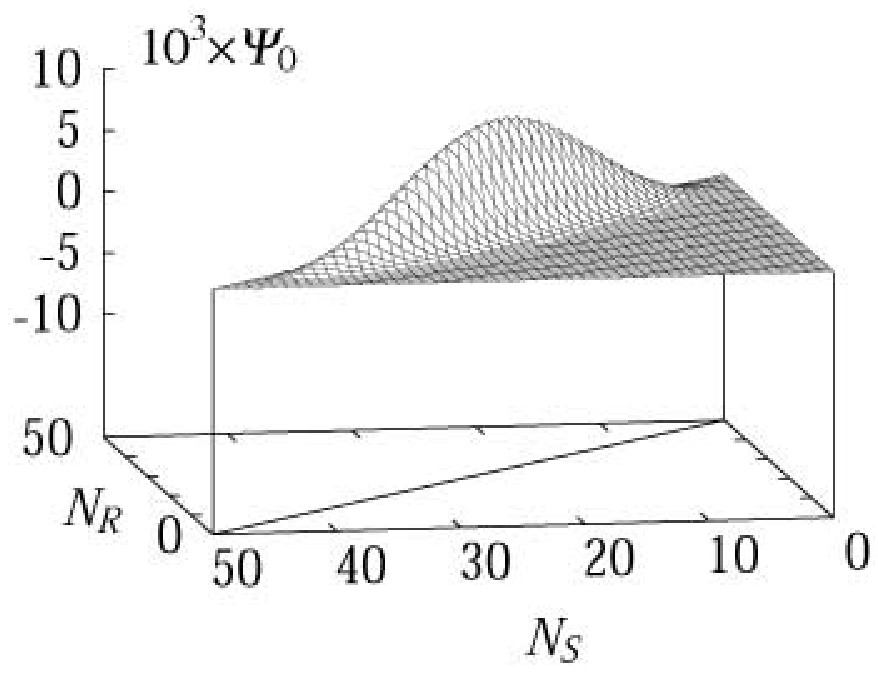}
\includegraphics[width=0.49\linewidth]{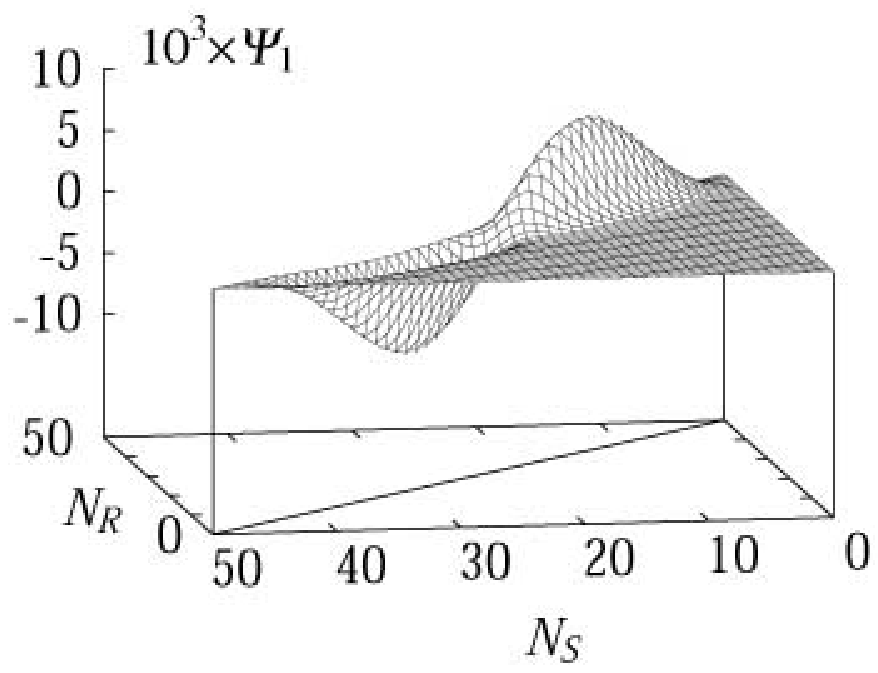}\\
(b) \hspace{8cm} (c)
\end{center} 
\caption{(a) Eigenvalues of the time-evolution matrix with a spontaneous and a
linear autocatalytic production and recycling; 
$\kappa_1=0.2k_0$, $\lambda=k_0$ and $k_2=0$.
(b) The eigenfunction $\Psi_0$ of the zero eigenvalue $\Lambda_0=0$, 
and (c) $\Psi_1$ of the second largest eigenvalue $\Lambda_1$ at $N=50$.
}
\label{fig4}
\end{figure}

\subsection{Quadratic autocatalysis}
In the case with a quadratic autocatalysis 
with parameters, 
$\kappa_2=4 \times 10^{-3}k_0$ and $\lambda=k_0, ~k_1=0$,
the second largest eigenvalue 
$\Lambda_1$ approaches 0 within a system size upto $N=$50,
whereas the third largest $\Lambda_2$ saturates to a finite value,
as shown in Fig. 5(a).
According to Eq.(\ref{eq31}), 
the critical size for the chiral symmetry breaking
is calculated to be $N_c=2(1+\lambda/k_0) (\kappa_2/k_0)^{-1/2} \approx 39.5$
in the present choice of the parameters. 
In a semi-logarithmic plot of the eigenvalues in Fig. 5(b), the
splitting of $-\Lambda_1$ and $-\Lambda_2$ is in fact taking place
close to this critical size $N_c$.
Near the maximum possible size $N=50$,
the size-depence of the eigenvalue 
$\Lambda_1$ looks like to be 
exponential as
$\Lambda_1=-2\times 10^3 \exp(-0.23N)$. 
Of course, the eigenvalue degeneracy seems just have commenced, and
the true asymptotics might be accomplished for much larger systems.
Therefore, the formula is only tentative.
Still, the emergence of non-ergodicity exists for sure,
and the phase transition takes place 
in a thermodynamic limit $N \rightarrow \infty$.

We know from the analysis in \S 2 that the final distribution
with this autocatalysis has a symmetric double peak structure.
That corresponds to the eigenfunction $\Psi_0$ 
for the eigenvalue $\Lambda_0=0$,
shown in Fig.5(c), and looks similar to the final probability distribution in Fig. 3(b) for a larger system $N=100$.
The eigenfunction $\Psi_1$ corresponding to $\Lambda_1$ is asymmetric, 
as shown in Fig. 5(d).
If one starts from an initial state with a small chirality, as in the
case of Fig.3(c), some component of $\Psi_1$ 
is mixed in the initial state so as to enhance one peak and 
suppress the other peak in  $\Psi_0=P_f$. 
As time passes, $i \ge 2$-components die off,
but the two components ($\Psi_0,\,\Psi_1$) remain 
over the duration $1/\Lambda_1$, 
and the asymptotic probability distribution
is single peaked at a chiral state, as shown in Fig.3(d).

When the recycling vanishes  ($\lambda\rightarrow 0$), 
not only one but many eigenvalues approach zero, 
irrespective of the system size, and
the system is completely non-ergodic.
The final state depends on the initial state.
We cannot apply the analysis employed here, and 
have to consider the problem in a completely different way
 in the next section.

\begin{figure}[h]
\begin{center} 
\includegraphics[width=0.49\linewidth]{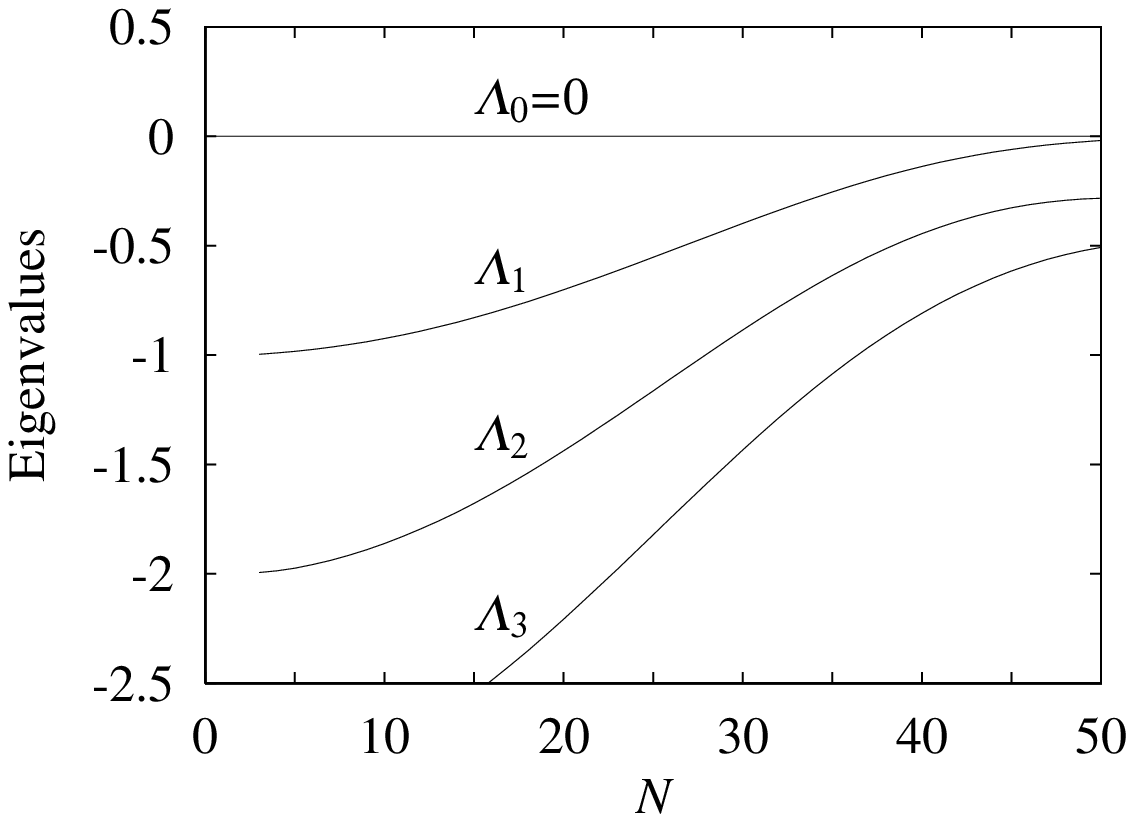}
\includegraphics[width=0.49\linewidth]{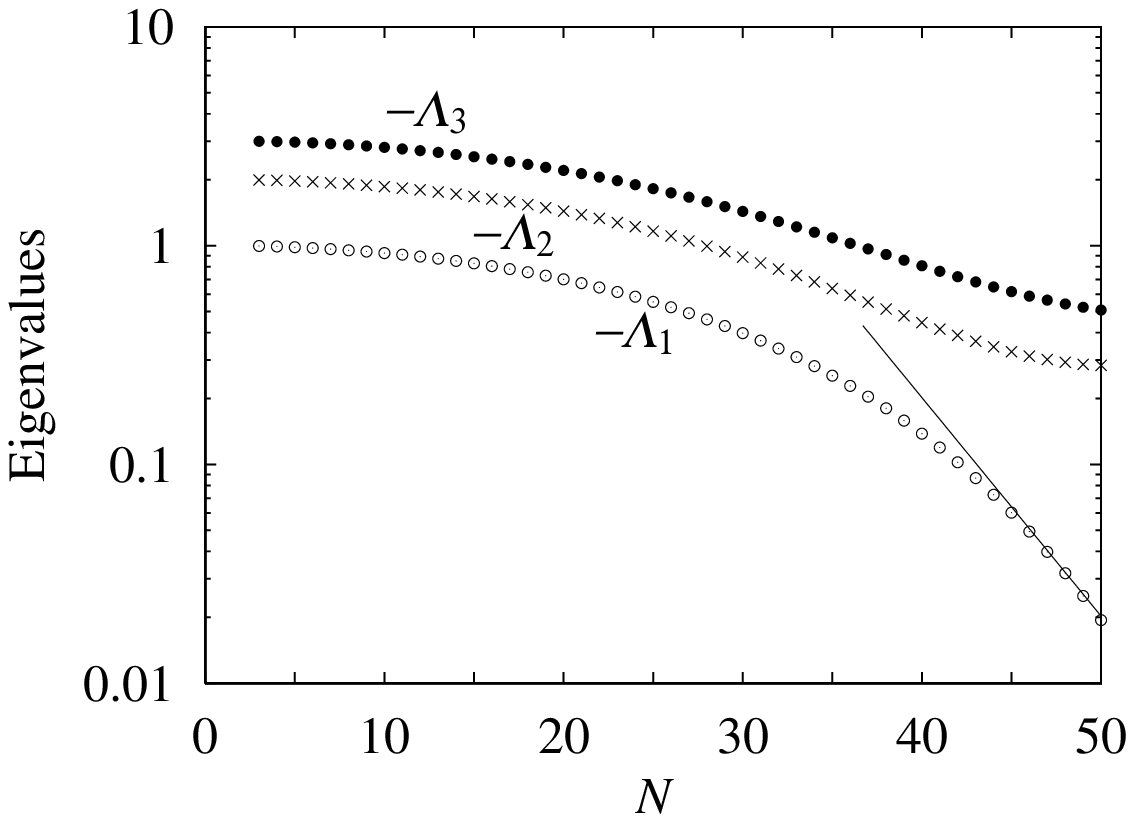}\\
(a) \hspace{8cm}(b)\\
\includegraphics[width=0.49\linewidth]{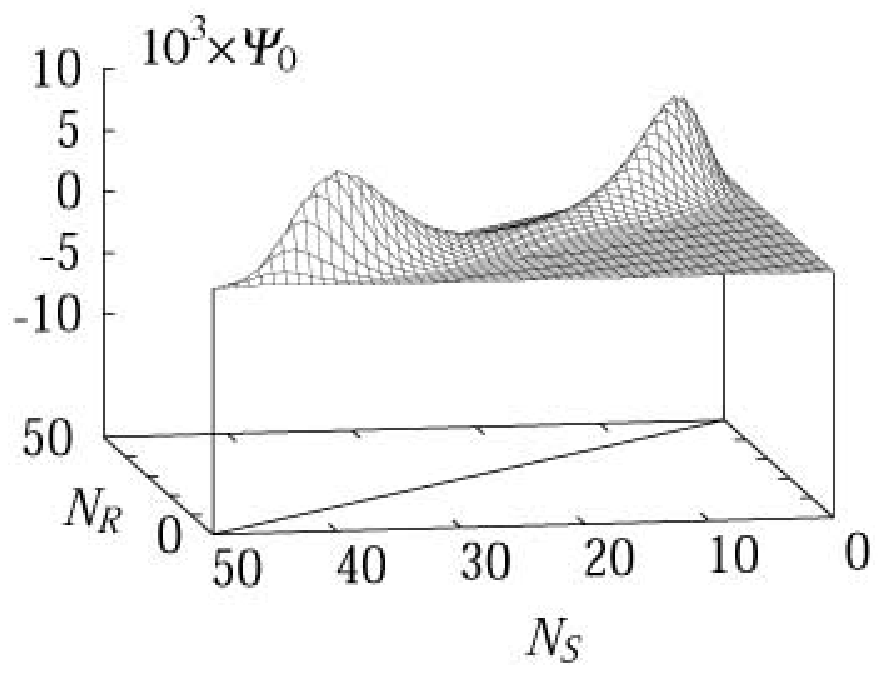}
\includegraphics[width=0.49\linewidth]{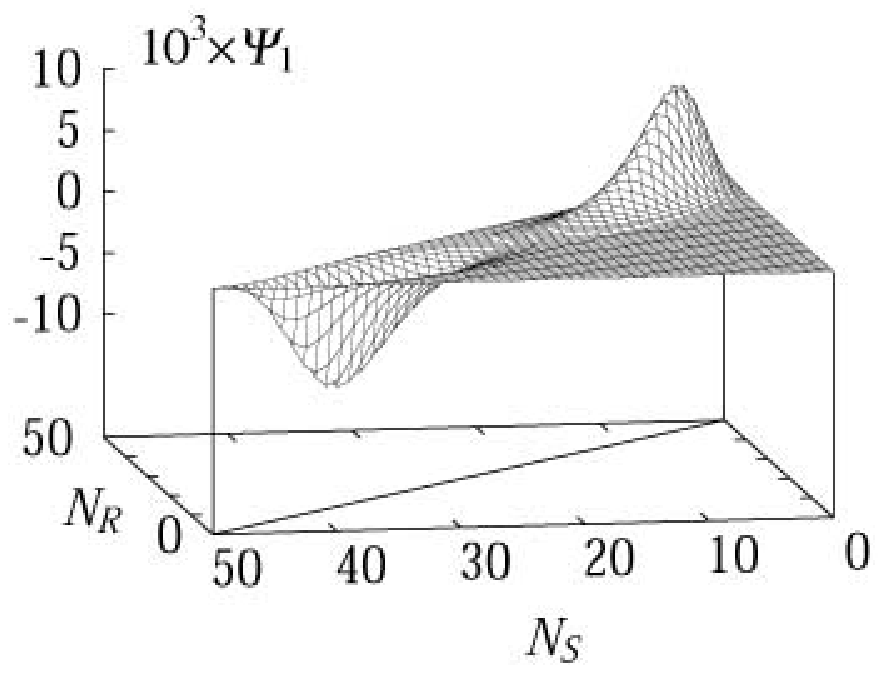}\\
(c) \hspace{8cm}(d)\\
\end{center} 
\caption{(a) A few largest eigenvalues 
of the time-evolution matrix with a spontaneous and a
quadratic autocatalytic production and recycling; 
$\kappa_2=4 \times 10^{-3} k_0,~\lambda=k_0$ and $k_1=0$. 
The second largest eigenvalue $\Lambda_1$ approaches 0
as the total number $N$ increases.
(b) Semi-logarithmic plot of the magnitude of a few largest eigenvalues.
The asymptotic behavior 
$\Lambda_1 \approx -2 \times 10^{3} \exp(-0.23 N)$
fits well the second largest eigenvalue $\Lambda_1$.
(c) The eigenfunction $\Psi_0$ with the zero eigenvalue $\Lambda_0=0$, and
(d) $\Psi_1$ of the second largest eigenvalue $\Lambda_1$ at $N=50$.
}
\label{fig5}
\end{figure}

\section{Directed Random Walk Model of Non-Recycling System}

So far there remains an apparent discrepancy between 
the final probability distribution 
for the nonrecycling system obtained by 
the time integration of the master equation and 
that obtained as a nonrecycling limit 
of the analytic result derived by the detailed balance consideration. 
For a spontaneous and for a linear autocatalytic cases,
these two analyses give the same results, 
whereas for a nonlinear autocatalytic case they are different.
In fact, in the nonrecycling case, the detailed balance condition 
cannot be imposed from the beginning,
since there is no back reaction to balance
the production process.
Therefore, one has to consider in a completely different way.

Here we propose a toy model which may be relevant to the stochastic
evolution of an autocatalytic system without recycling.
It is a random walk model on a square lattice $(N_R, N_S)$
in a triangular region $ 0 \le N_R, N_S, (N_R+N_S) \le N$.
A random walker can only make directed walks to the right or upwards:
$N_R \rightarrow N_R+1$ or $N_S \rightarrow N_S+1$.
The transition probability is Eq.(\ref{eq06}) to the right,
and the corresponding one to upwards.
This means that a walker on a site ($N_R, N_S)$ stays on the site 
for a waiting time interval
\begin{align}
\tau(N_R,N_S)=\frac{1}{
[2k_0+\kappa_1(N_R+N_S)+\kappa_2(N_R^2+N_S^2)]N_A}
\label{eq44}
\end{align}
with $N_A=N-N_R-N_S$, and then jumps
to the right and upwards with probabilities $p_r$ and $p_u$ given as 
\begin{align}
p_r(N_R,N_S)&=\frac{k_0+\kappa_1N_R+\kappa_2N_R^2}
{2k_0+\kappa_1(N_R+N_S)+\kappa_2(N_R^2+N_S^2)},
\nonumber \\
p_u(N_R,N_S)&=\frac{k_0+\kappa_1N_S+\kappa_2N_S^2}
{2k_0+\kappa_1(N_R+N_S)+\kappa_2(N_R^2+N_S^2)},
\label{eq45}
\end{align}
respectively.
The average position of the walker is expected to follow the evolution
similar to the rate equations Eq.(\ref{eq05}) with $\lambda=0$.
The remarkable point of this model is that 
though the waiting time $\tau(N_R,N_S)$ gets longer as $N_A$
decreases,
the rates of the transitions to
the right and upwards do not depend on $N_A$ nor vanish when $N_A=0$.
For a walker to reach on a line $N_R+N_S=N$,
it may take an exceedingly long time for an external observer, 
but it requires only $N$ steps for a walker himself.

Let the walker start from the origin $(N_R,N_S)=(0,0)$. 
After a time $t^{(1)}=\tau(0,0)=1/2k_0N$
he jumps to the right or upwards with the same probability, and lands
on a line $N_R+N_S=1$.
Wherever he is on a line $N_R+N_S=1$, he makes a second jump 
at a time $t^{(2)}=t^{(1)}+[(2k_0+\kappa_1+\kappa_2)(N-1)]^{-1}$
to the line $N_R+N_S=2$.
For a nonautocatalytic case with $\kappa_2 >0$, the time of the third 
( and the further) jump
depends on a position $N_R$ and $N_S$, but 
when $\kappa_2=0$ the walker jumps to the next line $N_R+N_S=3$ at a same time 
$t^{(3)}$. In fact, after the $n$-th jump, the walker corresponding to the 
linear autocatalysis is somewhere on a line 
$N_R+N_S=n$, and jumps to the next line $N_R+N_S=n+1$ at the same time
independent of the position on the line. 
In this case, we can calculate the probability evolution 
analytically.

\subsection{Directed Random Walker corresponding to Spontaneous and
Linear Autocatalysis}

A walker is making a random walk corresponding to the spontaneous production
and a linear autocatalysis with a finite $k_0$ and $\kappa_1$
 but without nonlinear autocatalysis ($\kappa_2=0$).
When he is on  a lattice site $(N_R,N_S)$, 
he has to wait for a next jump a time interval
$\tau (N_R,N_S)$ which depends only on 
the sum $N_R+N_S=n$ but not on $N_R$ and $N_S$ individually,
so that the waiting time can be denoted by $\tau_n$.
This sum $n$ also represents the number of jumps he has made, 
and totally $t^{(n)}=\sum_{i=0}^{n-1} \tau_i$ time has elapsed 
since he started from the origin $(N_R,N_S)=(0,0)$ 
till he reaches the present site.
After $n$ jumps, the walker 
is somewhere on a line $N_R+N_S=n$,
and we denote the probability as $P_{RW}(N_R,N_S; n)$.
It is non-zero on a line $N_R+N_S=n$ but zero anywhere else.
He should have come from the left or from the down.
Therefore, the probability should satisfy the relation
\begin{align}
&P_{RW}(N_R,N_S; n) 
\nonumber \\
&
=p_r(N_R-1,N_S)P_{RW}(N_R-1,N_S; n-1)
+p_u(N_R,N_S-1)P_{RW}(N_R,N_S-1; n-1)
\nonumber \\
&= \frac{
(k_0+\kappa_1 (N_R-1))P_{RW}(N_R-1,N_S; n-1)+
 (k_0+\kappa_1 (N_S-1))P_{RW}(N_R,N_S-1; n-1)}{
(2 k_0+\kappa_1 n)}
\nonumber \\
&=\frac{(N_R+N_S)!}{N_R!N_S!} \frac{
\prod_{i=0}^{N_R-1} (k_0+\kappa_1 i)
\prod_{j=0}^{N_S-1} (k_0+\kappa_1 j)}{
\prod_{l=0}^{N_R+N_S-1} (2 k_0+\kappa_1 l)}
P_{RW}(0,0; 0).
\label{eq46}
\end{align}
The combination factor represents the number of ways of arranging
the order of
right and upwards jumps among the $N_R+N_S=n$ total jumps from the initial state
 at the origin: $P_{RW}(0,0; 0)=1$.
After $n=N$ jumps, the walker reaches to the final line $N_R+N_S=N$,
and the distribution of the walker $P_{RW}(N_R,N_S)$ agrees with the 
final distribution (\ref{eq30}) discussed in \S 3.2
in the limit of $\lambda \rightarrow 0$, and naturally with the
result of time integration, shown in Fig. 1(d) and 2(d).

If the walker has started from an arbitrary site ($N_{R0},N_{S0})$,
he can be on a site $N_R \ge N_{R0}$ and $N_S \ge N_{S0}$ with a probability
\begin{align}
P_{RW}(N_R,N_S) 
=\frac{(N_R+N_S-N_{R0}-N_{S0})!}{(N_R-N_{R0})!(N_S-N_{S0})!} \frac{
\prod_{i=N_{R0}}^{N_R-1} (k_0+\kappa_1 i)
\prod_{j=N_{S0}}^{N_S-1} (k_0+\kappa_1 j)}{
\prod_{l=N_{R0}+N_{S0}}^{N_R+N_S-1} (2 k_0+\kappa_1 l)}.
\label{eq47}
\end{align}
It explains the asymmetric final probability distribution obtained by starting 
from a chiral initial state, shown in Fig.2(c).

\subsection{Directed Random Walker corresponding to Nonlinear Autocatalysis}
The same analysis cannot be performed on the time evolution of the
random walker corresponding to the nonlinear autocatalysis,
since his dwelling time on a site $(N_R,N_S)$ 
depends not only on the sum $N_R+N_S$ but also $N_R$ and $N_S$ 
individually.
But after the $n$th step, the walker is on a line $N_R+N_S=n$.
Only the arrival time depends on the path he took from the start at (0,0). 
If he comes on the site after the $n$th jump, he has come from the
left $(N_R-1,N_S)$ or from below $(N_R,N_S-1)$.
Therefore, the total probability $P_{RW}(N_R,N_S:n)$ 
that he passes the site $N_R+N_S=n$ satisfies the relation Eq.(\ref{eq46});
\begin{align}
&P_{RW}(N_R,N_S; n) =p_r(N_R-1,N_S)P_{RW}(N_R-1,N_S; n-1)
\nonumber \\
&+p_u(N_R,N_S-1)P_{RW}(N_R,N_S-1; n-1).
\label{eq48}
\end{align}
Because the denominators of $p_r(N_R-1,N_S)$ and $p_u(N_R,N_S-1)$
are not common, we cannot calculate the distribution analytically.
However, by starting from $P_{RW}(N_R=0,N_S=0:0)=1$,
we can calculate $P_{RW}(N_R,N_S:n)$ for any 
combinations of $N_R$ and $N_S$.
Since the random walker passes the particular site $(N_R,N_S)$ once
and for ever, and he stops on a line $N=N_R+N_S$ if one considers
the time evolution, the final probability distribution of the walker 
is given by $P_{RW}(N_R,N_S;N)$.

For the nonlinear autocatalytic system of a size $N=100$
 with reaction parameters $\kappa_2=10^{-3}k_0$ and $\kappa_1=\lambda=0$, 
we obtain the final probability distribution which has a single maximum at a racemic state,
in agreement with the result of time integration, as shown in Fig. 6(a).

\begin{figure}[h]
\begin{center} 
\includegraphics[width=0.35\linewidth]{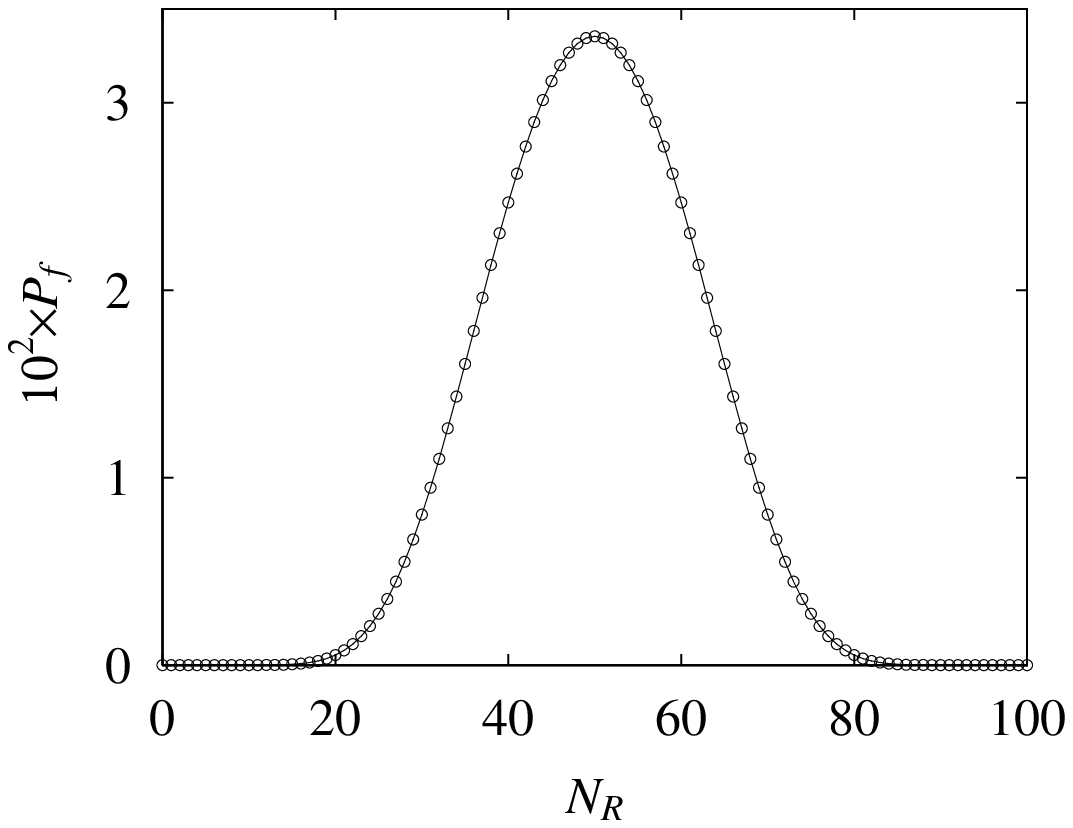}
\includegraphics[width=0.27\linewidth]{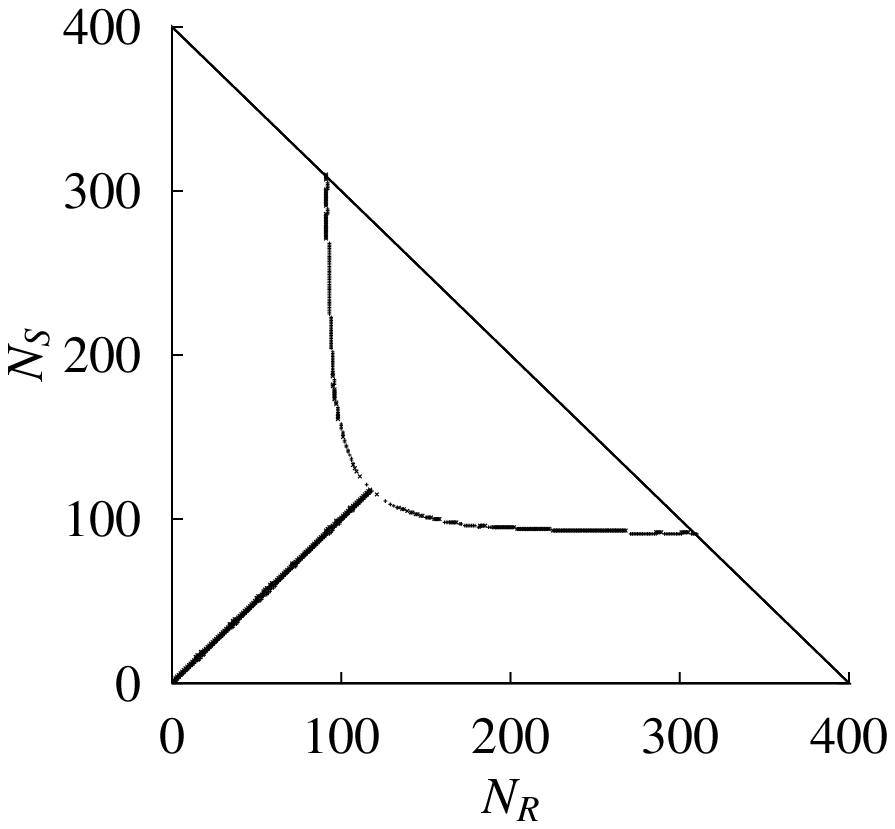}
\includegraphics[width=0.35\linewidth]{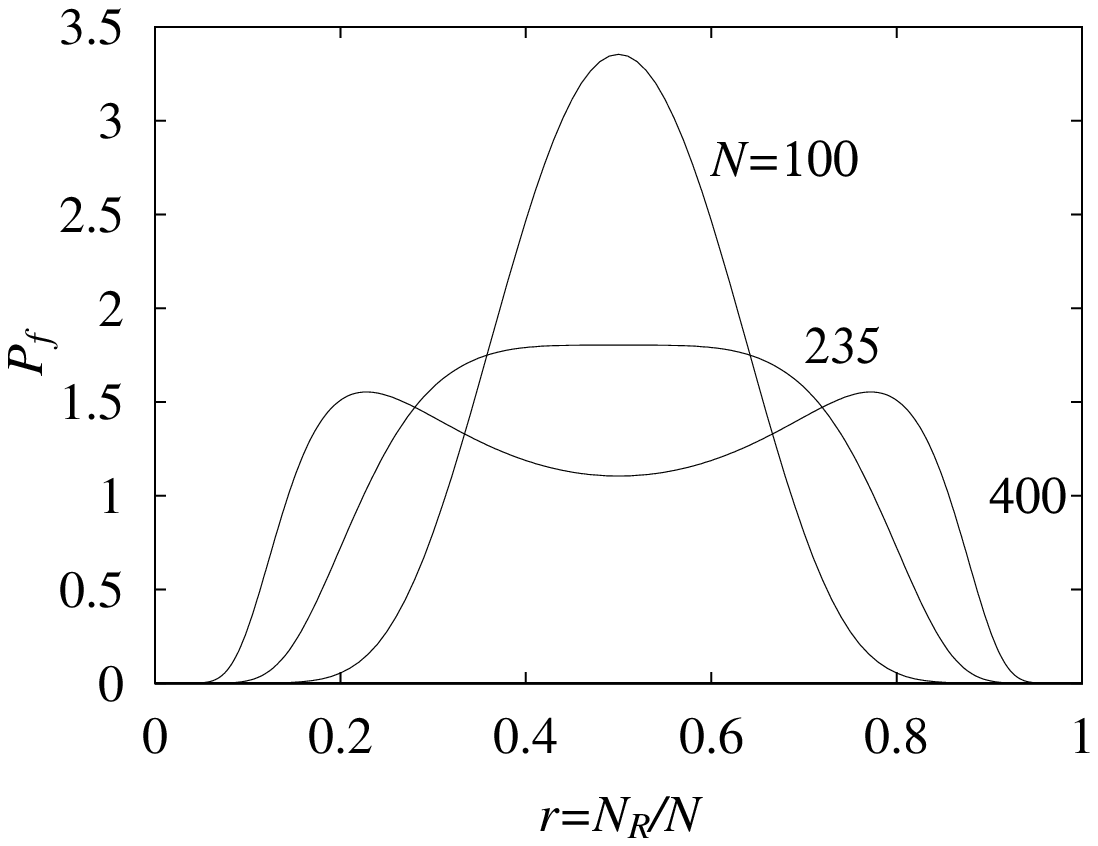}\\
(a) \hspace{4cm}(b)\hspace{4cm}(c)\\
\includegraphics[width=0.28\linewidth]{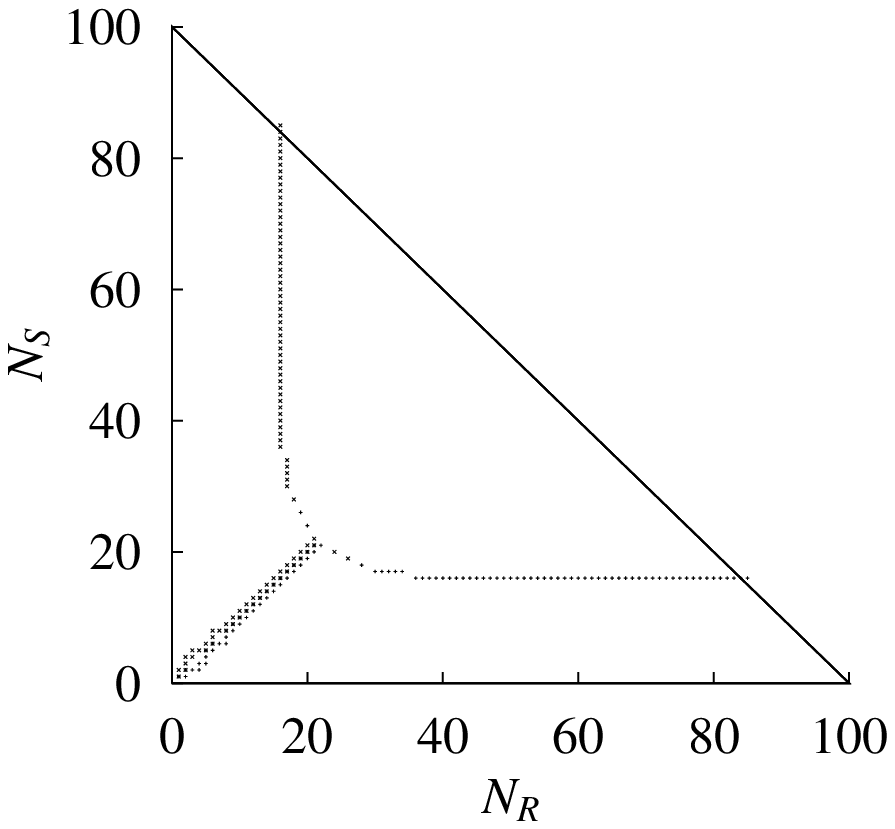}
\includegraphics[width=0.35\linewidth]{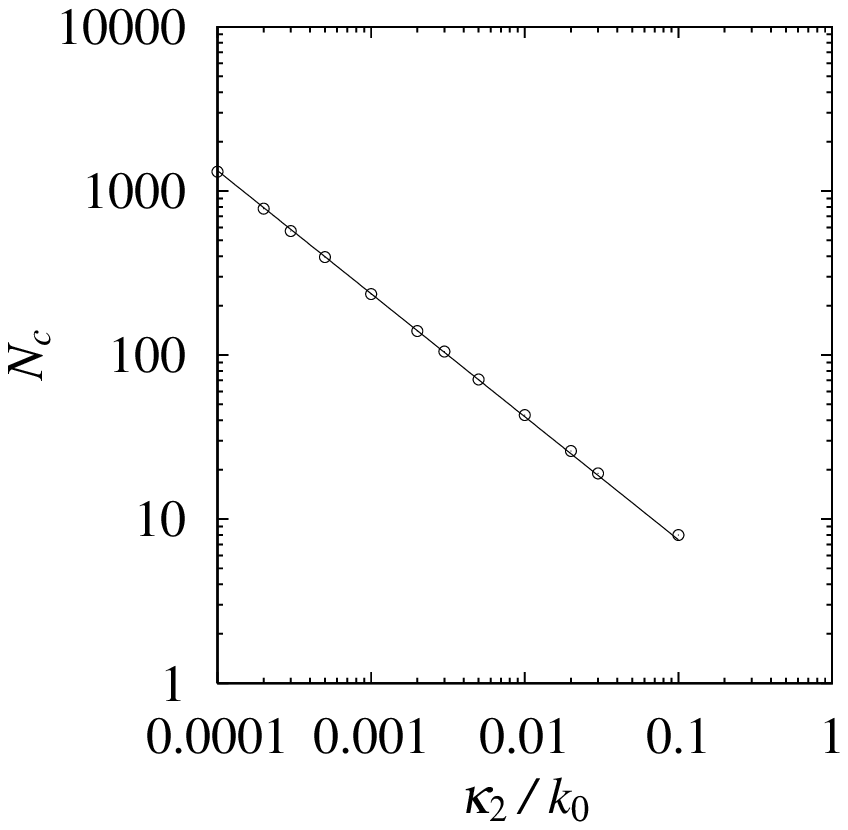}\\
(d) \hspace{8cm} (e)
\end{center} 
\caption{(a) Final probability distribution 
of a nonlinear autocatalytic system:
$N=100, ~\kappa_2=10^{-3}k_0, \kappa_1=\lambda=0$.
Symbols are the result by the time integration of the master equation
and a continuous curve represents numerical calculation of the 
directed random walk model.
(b) A trace of the probability maxima for $\kappa_2=10^{-3}$.
(c) Variation of the final probability distribution as a function of $N$.
(d) A trace of the probability maxima for $\kappa_2=10^{-2}$.
(e) Critical number $N_c$ for symmetry breaking at various strength
of quadratic autocatalytic rate constant $\kappa_2$.
The fit line is $N_c=1.33 (\kappa_2/k_0)^{-0.75}$.
}
\label{fig6}
\end{figure}

In Fig. 6(a) the probability distribution has a single peak, but
in fact, the profile depends on the total number of chemical reactants 
$N$.
The peak positions of the probability
distribution $P_{RW}$ are shown in Fig.6(b) up to a size $N=400$.
It has a single peak for small $N_R+N_S=N<N_c$,
whereas for large sizes $N>N_c$ it has symmetric double peaks.
The probability start to split to have a double-peak structure
around $N_c \approx 235$, and
for N=400 the probability has clear double peaks, as shown in Fig.6(c).
The critical value in terms of the concentration is
$k_{2c}c^2= \kappa_2 N_c^2 \approx 55.2k_0$, quite larger than the
value estimated from the $\lambda \rightarrow 0$ limit
value $k_{2c}c^2=4k_0$.
Another fact to be noticed in Fig. 6(b) is that 
the traces of the two peaks
are asymptotically parallel to two axes: 
$N_R$ at the peak position, for example, is fixed, 
but $N_S=N-N_R$ increases with $N$,
and so does
the magnitude of the ee order parameter $|\phi|=|N_R-N_S|/N$.

The critical number $N_c$ for double peaks actually
depends on $\kappa_2/k_0$. For example,
on increasing $\kappa_2$ to $\kappa_2=10^{-2}k_0$, 
the traces of probability maxima is scaled down, as shown in Fig. 6(d).
The critical size for the double-peak structure in this case
is $N_c \approx 43$,
and in terms of the rate constant $k_2$ its critical value is
$k_{2c}c^2 =\kappa_2 N_c^2= 18.5k_0$.
By varying $\kappa_2/k_0$ from $10^{-4}$ to $10^{-1}$,
one finds that the critical size $N_c$ depends on $\kappa_2/k_0$
as $N_c =1.33 (\kappa_2/k_0)^{-0.75}$, as shown in Fig. 6(e).
We will show later that the exponent $-3/4$ has a certain meaning.

When $\kappa_2$ is close to $k_0$ and $N_c$ is of the order unity,
the discreteness of the integer number comes into play in
the formula for $N_c$.
In fact, at $\kappa_2 \approx k_0$ the probability peak takes place 
on the boundary, $N_R=0$ and $N_S=0$.
A similar behavior is already found for the linearly autocatalytic case.
This represents a single mother scenario of the homochirality
by the autocatalytic reactions: 
When one starts from a completely achiral state without any
chiral ingredients initially, 
the first chiral species produced randomly by the spontaneous reaction
converts all the achiral substrate $A$ into her descendants before
the second mother of different chirality is born.

We now study the case when $\kappa_2$ is much smaller than $k_0$;
$\kappa_2 \ll k_0$.
Even in this case, the probability distribution has
 double peaks when the total number of reactants is large enough,
 as shown in Figs.6(b)-(d).
By starting from an achiral state without any chiral species,
the initial production is governed by the spontaneous production.
Only after $n$ steps of spontaneous production where 
\begin{align}
\kappa_2 n^2 \sim k_0,
\label{eq49}
\end{align}
there are enough chiral products for quadratic autocatalysis:
$n \sim \sqrt{k_0/\kappa_2} \gg 1$.
Here the probability distribution is Gaussian centered at a racemic state
$N_{R0}=N_{S0}=n/2$ with a width $n$. 
In terms of the difference $x_0=N_{R0}-N_{S0}$ at a step $n=N_{R0}+N_{S0}$, 
the probability is approximately given as
\begin{align}
P(x_0)dx_0 \propto e^{-x_0^2/2n} dx_0.
\label{eq50}
\end{align}
After the initial stage of $n$ steps,
 the quadratic autocatalysis comes into play.
The state with $(N_{R0},N_{S0})$ evolves deterministically 
according to the rate equations
of the quadratic autocatalysis. At the $N$-th step  a state 
reaches to  $(N_{R},N_{S})$ with $N_R+N_S=N$ and
related to the "initial" state $(N_{R0},N_{S0})$ as
\begin{align}
\frac{1}{N_R}-\frac{1}{N_S} =
\frac{1}{N_{R0}}-\frac{1}{N_{S0}} 
\label{eq51}
\end{align}
or in terms of the chirality $x=N_R-N_S$ as
\begin{align}
x_0=x \frac{n^2-x_0^2}{N^2-x^2} \approx
\frac{n^2}{N^2} x \left(1+\frac{N^2-n^2}{N^4} x^2 + \cdots\right).
\label{eq52}
\end{align}
The second approximation is valid close to the racemic state
with $|x_0/n|,~|x/N| \ll 1$.
The probability distribution of $x$ around a racemic state
$|x/N| \ll 1$ is obtained as
\begin{align}
P(x) dx = P(x_0)dx_0 \propto e^{-x_0^2/2n} \frac{dx_0}{dx} dx
\sim e^{-Ax^2/2+ {\cal O}(x^4)} dx
\label{eq53}
\end{align}
with the coefficient $A$ written as
\begin{align}
A= \frac{n^3}{N^4}-\frac{6}{N^2}
\label{eq54}
\end{align}
for $N \gg n \gg 1$.
The coefficient $A$ becomes negative for 
\begin{align}
N > N_c=\sqrt{\frac{n^3}{6}} \sim \Big( \frac{\kappa_2}{k_0} \Big)^{-3/4},
\label{eq55}
\end{align}
indicating that the probability distribution is minimum at $x=0$.
It means that the probability distribution has peaks
at somewhere different from the racemic point $x=0$. 
It should be a double peak structure because of the symmetry.
The power law dependence Eq.(\ref{eq55}) 
of the minimal number of chemical reactants $N$
on the ratio of rate constants with an exponent -3/4 
is in good agreement of the numerical data, shown in Fig. 6(e).

This mathematical analysis is qualitatively interpreted in the following
manner.
With a small $\kappa_2$, the chiral species are produced spontaneously
in the initial stage 
and the probability distribution has a racemic single peak.
After the numbers of chiral species $n$ are large enough as 
$n\sim \sqrt{k_0/\kappa_2}$,
the quadratic autocatalysis  increases the population of the
majority enantiomer more than that of the minority ones, 
and when there are totally $N$ chiral molecules,
the probability of the racemic state decreases with a fluctuation about
 $N^4/n^3$ larger than the Gaussian width $N$. 
Furthermore, the nonlinear relation
between the initial enantiomeric excess or $x_0$ to the final one $x$
changes the Jacobian. Therefore, after a sufficient steps of nonlinear
growth $N \sim (k_0/\kappa_2)^{-3/4}$, 
probability distribution takes a double peak structure.
The rate coefficient for the quadratic autocatalysis should be large as
$\kappa_2 N^{4/3} > k_0$ for the probability distribution to have a 
double-peak structure, but this condition is weaker 
than that for the single mother scenario; $\kappa_2 > k_0$.

\section{Summary and Discussions}

The probability distribution of the populations of chiral species in the reaction
of chiral molecule production is studied by means
of stochastic master equation.
With a recycling back reaction, the system relaxes to a
 unique final distribution, which can be obtained analytically 
 by assuming a detailed balance condition.
The final probability distribution has a racemic single peak for systems with
a spontaneous and a linear autocatalytic reactions.
With a quadratic autocatalysis, it has a double-peak
structure, indicating the chiral symmetry breaking.

Without recycling, the probability is shown to depend on the initial condition.
By starting from a complete achiral initial state without chiral ingredients,
the final distribution agrees with that predicted by the weak recycling limit 
obtained by the detailed balance condition 
in a spontaneous and a linear autocatalytic cases, but 
in a quadratic autocatalytic system detailed balance condition 
leads to erroneous final probability different from that obtained by the
time-integration of the master equation.
Without recycling, in fact, there is no reason that the detailed balance 
condition holds. Instead, we presented another directed random walk
model for the non-recycling system.
The final probability is explained in all the cases without recycling
by this model.
With the quadratic autocatalysis, the probability develops double peaks
as the rate of quadratic autocatalysis $k_2$ increases. 
Quadratic autocatalysis not only broadens the fluctuation 
produced initially by spontaneous production but also
increases the density of states away from the racemic state.
Therefore, if the rate coefficient $k_2$
is larger than that of the spontaneous production $k_0$ by a factor
about $V^2/N^{4/3}$, double peaks develop in the probability 
distribution, i.e. if the reaction takes place
in a small volume $V$ with a large number of reactants $N$.

In the limit of a very small rate of the spontaneous production
$k_0 \ll k_1/V$ or $k_0 \ll k_2/V^2$, 
the probability distribution
has double peaks where one of the enantiomer is missing.
This corresponds to the single mother, or chirality Eve, scenario
for the homochirality.
After the first chiral molecule is produced spontaneously and randomly,
all the available achiral substrate molecules are converted to the
first mother's type by linear or quadratic autocatalysis, before
the second mother of another enantiomeric type is born spontaneously.
This rare situation cannot be described by the rate equation but only
by the stochastic method.

As for the double peaks in the probability distribution found
in experiments of the Soai reaction, they are not sharp ones at the perfect homochirality.
Therefore, the single mother scenario is improbable, and
the quadratic autocatalysis with a moderate strength seems most
plausible.
However, if the single mother scenario is accompanied with some
imperfections as epimerization process \cite{plasson+04}
or erroneous catalysis,
peaks may shift to smaller ee values.
More careful study is necessary.


\end{document}